\documentclass[superscriptaddress,twocolumn]{revtex4}

\usepackage{amsmath}
\usepackage{amsfonts}
\usepackage{amssymb}
\usepackage{mathptmx}
\usepackage{graphicx}
\usepackage{epsfig}

\graphicspath{{Figures/}}

\begin{document}

\title{Intrinsic and extrinsic noise effects on the phase transition
of swarming systems and related network models}

\author{Jaime A. Pimentel} 
\affiliation{ 
Instituto de Ciencias F\'{\i}sicas, 
Universidad Nacional Aut\'onoma de M\'exico.  
Apartado Postal 48-3, Cuernavaca, Mor.62251, M\'exico}

\author{Maximino Aldana}
\email{max@fis.unam.mx}
\affiliation{
Instituto de Ciencias F\'{\i}sicas, 
Universidad Nacional Aut\'onoma de M\'exico.
Apartado Postal 48-3,  Cuernavaca, Mor.62251, M\'exico}

\author{Cristi\'an Huepe}
\affiliation{
614 N. Paulina St., Chicago, IL 60622-6062,
USA}

\author{Hern\'an Larralde}
\affiliation{
Instituto de Ciencias F\'{\i}sicas, 
Universidad Nacional Aut\'onoma de M\'exico.
Apartado Postal 48-3,  Cuernavaca, Mor.62251, M\'exico}

\begin{abstract}
We analyze order-disorder phase transitions driven by noise that occur
in two kinds of network models closely related to the self-propelled
model proposed by Vicsek et. al. to describe the collective motion of
groups of organisms [\emph{Phys. Rev. Lett.} {\bf 75}:1226
(1995)]. Two different types of noise, which we call intrinsic and
extrinsic, are considered. The intrinsic noise, the one used by Vicsek
et. al. in their original work, is related to the decision mechanism
through which the particles update their positions. In contrast, the
extrinsic noise, later introduced by Gr\'egoire and Chat\'e
[\emph{Phys. Rev. Lett.} {\bf 92}:025702 (2004)], affects the signal
that the particles receive from the environment. The network models
presented here can be considered as the mean-field representation of
the self-propelled model. We show analytically and numerically that,
for these two network models, the phase transitions driven by the
intrinsic noise are continuous, whereas the extrinsic noise produces
discontinuous phase transitions. This is true even for the small-world
topology, which induces strong spatial correlations between the
network elements. We also analyze the case where both types of noise
are present simultaneously. In this situation, the phase transition
can be continuous or discontinuous depending upon the amplitude of
each type of noise.
\end{abstract}

\maketitle

\section{Introduction}

In spite of many efforts, only a limited understanding has been
achieved regarding the emergence of collective order in
non-equilibrium systems. While these systems often present features
analogous to the those found in equilibrium, such as phase-transitions
and long-range correlations, it is not clear how to use the powerful
tools available in equilibrium statistical mechanics to analyze their
properties. To overcome this problem, it may be useful to consider
cases where simple qualitative characteristics (e.g. the existence and
order of a phase transition) are common to different non-equilibrium
systems with similar properties.

A class of non-equilibrium systems that has sparked increasing
interest in recent years is given by models of groups of swarming
agents \cite{Grunbaum,Vicsek,Czirok1,Czirok2,Nagy,Topaz1}. These are
used to describe the collective behavior of self-propelled agents such
as schools of fish, flocks of birds, or herds of quadrupeds
\cite{Parrish,Couzin1,Couzin2}. Even the simplest of these models
displays large-scale organized structures, in which agents separated by
distances much larger than their interaction ranges can coordinate and
swarm in the same direction. If noise is added to the system, this
ordered state is destroyed as the noise level increases. When the
noise reaches a critical value, the system undergoes a phase
transition to a disordered state where agents move in random directions
\cite{Vicsek, Czirok1}. This phase transition has been quite
thoroughly analyzed through numerical simulations. However, the lack
of a systematic theoretical approach to non-equilibrium systems has
hindered a proper characterization of the order-disorder phase
transition, and there are still doubts about its basic features even
in this simplest of cases \cite{Chate,OurPRL,ChateComment}.

In this paper we are particularly interested in how this phase
transition might be affected by the way in which the noise is
introduced in the system. Two different types of noise, which we will
call \emph{extrinsic} and \emph{intrinsic}, have recently been
considered in models of swarming \cite{Vicsek,Chate}. In these models,
at every time step each particle receives a signal from its neighbors
that tells the particle in which direction to move next. The extrinsic
noise consists in that the signal received by the particle is blurred
(because, say, the environment is not completely transparent and the
particle cannot see its neighbors very well). As a consequence, the
particle may move in a different direction to the one dictated by the
neighbors. In contrast, in the intrinsic noise case each particle
receives the signal sent by the neighbors perfectly, but then it may
``decide'' to do something else and move in a different
direction. Thus, the extrinsic noise can be thought of as produced by
a blurry environment, whereas the intrinsic noise comes from the
``free will'' of the particles, so to speak; namely, from the
uncertainty in the particle's decision mechanism. In either case, of
course, the net result is that, at every time step, the particle may
move in a direction that departs from the one dictated by the
neighbors.

It has been pointed out that these two distinct types of noise,
extrinsic and intrinsic, can produce very different order-disorder
phase transitions \cite{OurPRL}. This can been shown analytically
using a network approach in which the elements, instead of interacting
with the neighbors in a physical space, interact with any element that
is linked to them through a network connection. This kind of
description has been used to model a large range of dynamics, such as
the traffic between Internet websites or servers, the evolution of an
epidemic outbreak, the mechanisms triggered by gene expressions in the
cell, or the activity of the brain
\cite{Barabasi1,Strogatz,Watts3,Watts4,ReviewBarabasi,ReviewNewman}. In
the context of swarming systems, the network approach is equivalent to
a mean-field theory in which correlations between the particles are
not taken into account. However, this approach has the virtue that it
allows us to separate clearly the dynamical interaction rule
that determines the dynamical state of the particles, from the
topology of the underlying network that develops in time and space and
dictates who interacts with who. Therefore, under the network approach
it is possible to focus on the effects that the two different types of
noise have on the dynamics of the system. 

Even when the network approach leaves aside some important aspects of
the dynamics of swarming systems (such as correlations in space and
time), some appealing analogies can be established between the
swarming and network systems. Indeed, in the simplest swarming models,
the dynamics is defined by giving to each agent a steering rule that
uses the velocities of all agents in its vicinity as an input to
compute its own velocity for the next time-step. This algorithm can be
associated to a dynamics on a switching network that links at every
time-step all agents that are within the interaction range of each
other. In this context, the network is simply a representation of the
spatial dynamics of the system. However, it has been shown that this
analogy can be pushed further successfully and that a static network
with long-range connections can capture some of the main qualitative
behaviors of simple swarming models \cite{Huepe1,Aldana1,OurPRL}. 

In this paper, we compare the properties of the phase transitions and
dynamical mappings of two kinds of network models to further explore
the analogies described above. We consider models that incorporate
three of the main aspects of the interaction between the particles in
swarms: an average input signal from the neighbors, noise, and, in
some sense, extremely long-range interactions.  In the first kind of
model the elements of the network can acquire only two states, +1 and
-1; whereas in the second, the elements are represented by 2D vectors
whose angles take any value between 0 and $2\pi$. We find that
swarming systems and their network counterparts indeed present
qualitatively similar behaviors depending on whether the noise is
intrinsic or extrinsic.  We also determine numerically that the same
qualitative features arise when the particles are placed on a
small-world network, and we extend our results to the case in which
the network models are subject to both types of noise.

The paper is organized as follows. In Sec.~\ref{sec:vicsek} we present
the model introduced by Vicsek and his group to describe the emergence
of order in swarming systems. In particular, we focus our attention on
how the phase transition seems to change when the noise changes from
intrinsic to extrinsic. In Section~\ref{sec:voter} we present a
majority voter model on a network, which is reminiscent of the Ising
model with discrete internal degrees of freedom. This model is simple
enough as to be treated analytically, at least for the case of
homogeneous random network topologies for which we
show analytically that the two types of noise indeed
produce two different types of phase transition.  In
Sec.~\ref{sec:vectorial} we introduce another network model in which
the internal degrees of freedom are continuous (2D vectors). This
model can be treated analytically in the limit of infinite network
connectivity. However, these results and extensive numerical
simulations clearly indicate that the two types of noise again produce
two different phase transitions, which are analogous to the ones
observed in the majority voter model and in the self-propelled
model. In Sec.~\ref{sec:mean-field} we discuss the mean-field
assumptions conveyed in the two network models and how they relate to
the self-propelled model. We also show that the nature of the phase
transition produced by each type of noise does not change when the
small-world topology is implemented, which produces strong spatial
correlations between the network elements.  Finally, in
Section~\ref{sec:conclusions} we summarize our results.

\section{The Vicsek model}
\label{sec:vicsek}

Arguably, the simplest model to describe the collective motion of a
group of organisms was proposed by Vicsek and his collaborators
\cite{Vicsek}. In this model, $N$ particles move within a 2D box of
sides $L$ with periodic boundary conditions. The particles are
characterized by their positions, $\vec{x}_1(t),\dots,\vec{x}_N(t)$,
and their velocities $\vec{v}_1(t)=v
e^{i\theta_1(t)},\dots,\vec{v}_N(t)=v e^{i\theta_N(t)}$, (represented
here as complex numbers). All the particles move with the same speed
$v$. However, the direction of motion $\theta_n(t)$ of each particle
changes in time according to a rule that captures in a qualitative way
the interactions between organisms in a flock. The basic idea is that
each particle moves in the average direction of motion of the
particles surrounding it, \emph{plus some noise}. Two interaction
rules have been considered in the literature, which differ in the way
the noise is introduced into the system. To state these rules
mathematically, we need some definitions. Let ${\mathcal R}_n(r)$ be
the circular vicinity of radius $r$ centered at $\vec{x}_n(t)$, and
$K_n(t)$ be the number of particles whose positions are within
${\mathcal R}_n(r)$ at time $t$. We will denote as $\vec{U}_n(t)$ the
average velocity of the particles which at time $t$ are within the
vicinity ${\mathcal R}_n(r)$, namely
\begin{equation}
\vec{U}_n(t) = \frac{1}{K_n(t)} \sum_{\{j\,:\,\vec{x}_j(t)\in{\mathcal R}_n(r)\}}\vec{v}_j(t).
\end{equation}

For reasons that will be clear later, we will call $\vec{U}_n(t)$
\emph{the input signal} received by the $n$-th particle
$\vec{v}_n$. With the above definitions, the interaction rule
originally proposed by Vicsek \emph{et al.} can be written as
\begin{subequations}
\label{eq:SPMIN}
\begin{eqnarray}
\theta_n(t+\Delta t) &=& \mbox{Angle}\left[\vec{U}_n(t)\right]
+ \eta\xi_n(t), \label{eq:vicsek}\\
\vec{v}_n(t+\Delta t) &=& ve^{i\theta_n(t+\Delta t)},\\
\vec{x}_n(t+\Delta t) &=& \vec{x}_n(t) + \vec{v}_n(t+\Delta t)\Delta t,\label{eq:kinetic1}
\end{eqnarray}
\end{subequations}
where $\xi_n(t)$ is a random variable uniformly distributed in the
interval $[-\pi,\pi]$, and the \emph{noise amplitude} $\eta$ is a
parameter taking a constant value in $[0,1]$. The ``Angle'' function
is defined in such a way that if $\vec{u} = ue^{i\theta}$, then
Angle$[\vec{u}] = \theta$. Note that in this case the direction of the
neighbors' average velocity $\vec{U}_n(t)$ (the input signal) is
computed first and then the noise is added to this direction. We will
refer to the interaction rule given in Eqs.~\eqref{eq:SPMIN} as the
\emph{self-propelled model with intrinsic noise} (SPMIN), and to the
term $\eta\xi_n(t)$ as the \emph{intrinsic noise}.

The second interaction rule, proposed by Gr\'egoire and Chat\'e in
Ref.~\cite{Chate}, is given by
\begin{subequations}
\label{eq:SPMEN}
\begin{eqnarray}
\theta_n(t+\Delta t) &=& \mbox{Angle}\left[\vec{U}_n(t)+ \eta e^{i\xi_n(t)}\right],
\label{eq:chate}\\
\vec{v}_n(t+\Delta t) &=& ve^{i\theta_n(t+\Delta t)},\\
\vec{x}_n(t+\Delta t) &=& \vec{x}_n(t) + \vec{v}_n(t+\Delta t)\Delta t,\label{eq:kinetic2}
\end{eqnarray}
\end{subequations}
where $\xi_n(t)$ and $\eta$ have the same meaning as in
Eq.~(\ref{eq:vicsek}). In this case, a random vector of constant
length $\eta$ and random orientation $\xi_n(t)$ is added to the
neighbors' average velocity $\vec{U}_n(t)$, and then the direction of
the resultant vector is computed. We will refer to the model given in
Eqs.~\eqref{eq:SPMEN} as the \emph{self-propelled model with extrinsic
noise} (SPMEN), and to the term $\eta e^{i\xi_n(t)}$ as the
\emph{extrinsic noise}.

One might think that the two interaction rules (\ref{eq:vicsek}) and
(\ref{eq:chate}) are more or less equivalent and should produce
qualitatively similar dynamical behaviors. However, as mentioned in
the introduction, there is a clear physical difference between these
two ways of adding noise to the system. In the SPMIN the uncertainty
induced by the noise is in the decision mechanism (the Angle
function), but not in the input signal that each particle receives. In
contrast, one could say that in the SPMEN the particles are
``short-sighted'' and do not see well the signal sent by the neighbors, 
which is what causes the uncertainty in this case.
Thus, one cannot expect, \emph{a priori}, the onset of collective order to
be the same in both the SPMIN and the SPMEN, for the way in which the
noise is introduced in both cases is clearly different, not only
algorithmically, but also physically.

To measure the amount of order in the system we define the
instantaneous value of the order parameter $\psi(t)$ as
\begin{equation}
 \psi(t) = \left|\frac{1}{vN}\sum_{n=1}^N \vec{v}_n(t)\right| =
 v^{-1}\left|\langle \vec{U}(t)\rangle\right|,
\label{eq:orderParameter}
\end{equation}
where $\langle \vec{U}(t)\rangle=\frac{1}{ N}\sum_{n=1}^N
\vec{v}_n(t)$ is the average velocity of the entire system at time
$t$.  Thus, if $\psi(t)\approx0$ the particles move in random
uncorrelated directions, whereas if $\psi(t)\approx1$, all the
particles are aligned and move in the same direction. In the limit
$t\to\infty$, the order parameter $\psi(t)$ reaches a constant value
$\psi=\lim_{t\to\infty}\psi(t)$ that characterizes the steady state
behavior of the system \footnote{In the numerical simulations, the
stationary value $\psi$ of the order parameter is computed as
$\psi=1/T\int_0^T \psi(t)dt$ for large $T$.}.

\begin{figure}[t]
\scalebox{0.95}{\includegraphics{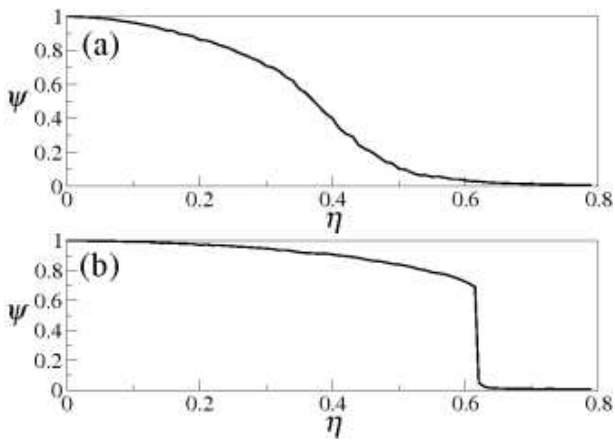}}
\caption{Phase transition in the self-propelled model with (a)
intrinsic noise and (b) extrinsic noise. In the first case the phase
transition seems to be continuous, whereas in the second case it is
clearly discontinuous. The numerical simulations were performed using
systems with $N=20000$, $L=32$, $r=0.4$ and $v=0.05$.}
\label{fig:phaseVicsek}
\end{figure}

In their original work, Vicsek and his group noted the existence of a
phase transition from ordered states ($\psi(t)>0$) to disordered
states ($\psi(t)\approx0$) as the value of the noise intensity $\eta$
increases. We illustrate this phase transition for the SPMIN in
Fig.~\ref{fig:phaseVicsek}(a), which was obtained numerically for a
system with $N=20000$ particles within a box of sides $L=32$. The
interaction radius and particle speed used in the numerical simulation
are $r=0.4$ and $v=0.05$, respectively. On the other hand,
Fig.~\ref{fig:phaseVicsek}(b) shows the phase transition for the SPMEN
with exactly the same parameters as in the previous case. As it can be
seen from Fig.~\ref{fig:phaseVicsek}, the phase transition looks
continuous (second order) for the SPMIN, whereas it is clearly
discontinuous (first order) for the SPMEN.

Numerical simulations performed for larger systems than the one used
in Fig.~\ref{fig:phaseVicsek} also seem to indicate that the phase
transition exhibited in the SPMIN is continuous
\cite{Czirok1,Czirok2,Nagy}. Nonetheless, based on numerical
simulations, the authors of Ref.~\cite{Chate} have pointed out that
the phase transition may be discontinuous regardless of the type of
noise, and that the apparent continuity of the phase transition in the
SPMIN is due to strong finite-size effects.  For some reason, these
finite size effects are not so strong in the SPMEN, in which the phase
transition is clearly discontinuous. Due to a lack of a general
mathematical formalism to analyze the self-propelled model, (either
with intrinsic or extrinsic noise), the way in which each type of
noise affects the dynamics of the system remains unknown. In what
follows we present two simplified versions of the Vicsek model that
can be solved analytically for the two types of noise. We show that in
these simpler models, the phase transition is continuous for the
intrinsic noise and discontinuous for the extrinsic noise.

\section{The majority voter model}
\label{sec:voter}

The first network model that we consider consists of a set of $N$
binary variables, $v_1,v_2,\dots,v_N$, each acquiring the values +1 or
-1. The value of each $v_n$ changes in time and is determined by a set
of $k_n$ other elements, ${\mathcal
I}_n=\{v_{n_1},v_{n_2},\dots,v_{n_{k_n}}\}$, which we will call the
\emph{inputs} of $v_n$. Thus, at every time step every element $v_n$
receives a signal from its set of inputs ${\mathcal I}_n$ and updates
its value according to that signal. Different ways of assigning the
inputs to each element lead to different network topologies. In this
section we focus on the homogeneous random topology characterized by
the following two properties: 
\begin{itemize}
\item All the elements have the same number of inputs $K$, namely,
$k_n=K$ for all $n$.
\item The $K$ inputs of each element are chosen randomly from anywhere
in the system.
\end{itemize}
Note that this is a directed network, for if $v_n$ is an input to
$v_m$, then $v_m$ is not necessarily an input to $v_n$.  We can think
of this system as a society of $N$ individuals in which every
individual $v_n$ can have two opinions (+1 and -1) about an
issue. Each individual's opinion is influenced by its $K$ friends
(inputs), who are randomly chosen among the $N$ individuals in the
society. Generally, each individual will tend to be of the same
opinion as the majority of its friends, but with a given probability
it can have the opposite opinion. This probability of having an
opinion opposite to that of the majority can be considered as a
``temperature'' that introduces noise in the dynamics of the
system. Here we consider two ways of introducing this noise, which are
analogous to the intrinsic and extrinsic noise in the self-propelled
model.

We define the input signal $U_n(t)$ influencing element $v_n(t)$ as 
\begin{equation}
U_n(t) = \frac{1}{K}\sum_{j=1}^K v_{n_j}(t).
\label{eq:inputSignalVoter}
\end{equation}
This is just the average opinion of the inputs of $v_n$. With this
definition, the dynamics of the system with \emph{intrinsic noise} is
given by the simultaneous updating of the network elements according
to the rule
\begin{equation}
v_n(t+1) = \left\{
\begin{array}{rll}
\mbox{Sign}\left[ U_n(t)\right] & \mbox{ with prob. } & 1-\eta \\
 & & \\
- \mbox{Sign}\left[ U_n(t)\right] & \mbox{ with prob. } & \eta \\
\end{array}
\right.
\label{eq:vicsekVoter1}
\end{equation}
where Sign$[U_n] = -1$ if $U_n<0$, and Sign$[U_n] = 1$ if $U_n>0$. If
$U_n=0$ then we choose for $v_n(t+1)$ either +1 or -1 with equal
probability. The \emph{noise amplitude} $\eta$ is a constant parameter
in the interval $[0,1/2]$ that represents the probability for each
individual to go against the majoritarian opinion. The above
interaction rule can also be written in a simpler form as
\begin{equation}
v_n(t+1) =\mbox{Sign}\left[\mbox{Sign}\left[U_n(t)\right] +
\frac{\xi_n(t)}{1-\eta}\right],
\label{eq:vicsekVoter2}
\end{equation}
where $\xi_n(t)$ is a random variable uniformly distributed in the
interval $[-1,1]$. From the above expression it is clear that this way
of introducing the noise is equivalent to the intrinsic noise of
Eq.~(\ref{eq:vicsek}), since the noise is added after the Sign
function has been applied to the input signal $U_n(t)$.  (Since $v_n$
can only take the values +1 or -1, the Sign function has to be applied
again.)

The other way of introducing the noise, which is equivalent to the
extrinsic noise in Eq.~(\ref{eq:chate}), is given by the interaction
rule
\begin{equation}
v_n(t+1) =\mbox{Sign}\left[U_n(t) + 4\eta\xi_n(t)\right],
\label{eq:chateVoter}
\end{equation}
where now the noise is directly added to the input signal $U_n(t)$ and
then the Sign function is evaluated. $\xi_n(t)$ and $\eta$ have the
same meaning as in Eq. ~(\ref{eq:vicsekVoter2}). (The factor 4 in the
preceding equation is just to guarantee that the phase transition in
this case occurs within the interval $\eta\in[0,1/2]$.)

\begin{figure}[t]
 \scalebox{0.95}{\includegraphics{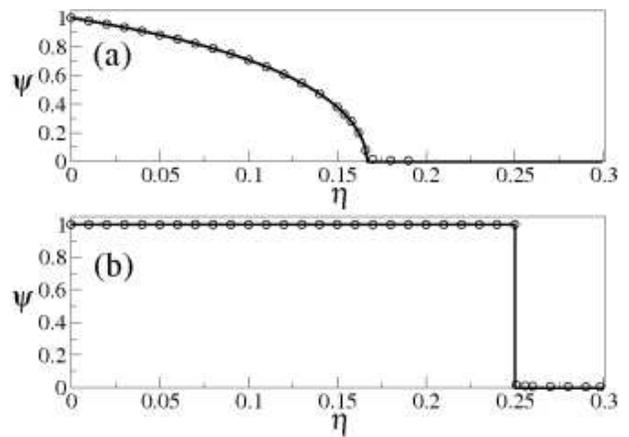}}
\caption{Phase transition in the majority voter model with (a)
intrinsic noise and (b) extrinsic noise. In the first case the phase
transition is continuous, whereas in the second case it is
discontinuous. The symbols represent data obtained from numerical
simulations using systems with $N=10^5$ and $K=3$. The solid lines
correspond to the mean-field prediction.}
\label{fig:phaseVoterK3}
\end{figure}

For both types of noise, intrinsic and extrinsic, the majority voter
model exhibits a phase transition from ordered to disordered
states. However, the nature of this phase transition (i.e. whether
continuous or discontinuous) depends on the type of noise.  To see
that this is indeed the case, we define the order parameter $\psi(t)$
for the majority voter model as
\begin{equation}
\psi(t) = \frac{1}{N}\sum_{n=1}^N v_n(t).
\label{eq:orderParameterVoter}
\end{equation}
In the limit $t\to\infty$, the order parameter $\psi(t)$ reaches a
stationary value $\psi$ that depends on the noise intensity
$\eta$. Fig.~\ref{fig:phaseVoterK3}(a) shows $|\psi|$ as a function of
$\eta$ for the intrinsic noise case [Eq.~\eqref{eq:vicsekVoter2}], in
a system with $N=10^5$ and $K=3$. It is apparent that in this case the
phase transition is continuous. This result is consistent with the
behavior of the SPMIN reported in
Fig.~\ref{fig:phaseVicsek}(a). Contrary to the above, the phase
transition for the majority voter model with extrinsic noise
[Eq.~\eqref{eq:chateVoter}] is discontinuous, as is shown in
Fig.~\ref{fig:phaseVoterK3}(b), which is also consistent with the
behavior observed in Fig.~\ref{fig:phaseVicsek}(b) for the
SPMEN. Thus, changing the way in which the noise is introduced in the
voter model also changes drastically the nature of the phase
transition.

The majority voter model is simple enough to be treated
analytically. We can even generalize the model to incorporate the two
types of noise simultaneously. In this generalization, the value of
each element $v_n$ is updated according to the dynamical rule
\begin{equation}
 v_n(t+1) =\mbox{Sign}\left[\mbox{Sign}\left[U_n(t) + 4\eta_1\xi_n(t)\right] +
\frac{\zeta_n(t)}{1-\eta_2}\right],
\label{eq:voterGeneral}
\end{equation}
where $\xi_n$ and $\zeta_n$ are independent random variables uniformly
distributed in the interval $[-1,1]$, and $\eta_1$ and $\eta_2$ are
constant parameters taking values in the interval $[0,1/2]$. Thus, if
$\eta_1=0$ and $\eta_2\neq0$, only the intrinsic noise is present,
whereas if $\eta_1\neq0$ and $\eta_2=0$ only the extrinsic noise is
present. Intermediate cases are obtained if both $\eta_1$ and $\eta_2$
are different from zero. In what follows we consider separately the
case in which $K$ is finite, and the case in which $K$ is infinite.

\subsection{Case 1: $K < \infty$}

In Appendix~A we present a mean-field calculation showing that, when
the network connectivity $K$ is finite, the order parameter $\psi(t)$
satisfies the dynamical mapping
\begin{subequations}\label{eqs:MappingVoter}
\begin{equation}
\psi(t+1) = M\left(\psi(t)\right),
\label{eq:mappingVoter1}
\end{equation}
where
\begin{equation}
M(\psi) = (1-2\eta_2)\sum_{m=1}^K\beta_m^K(\eta_1)\psi^m,
\label{eq:M-psi-Voter}
\end{equation}
and the coefficients $\beta_m^K(\eta_1)$ are given by
\begin{eqnarray}
\beta_m^K(\eta_1)&=&\binom{K}{m}\frac{(-i)^{m-1}}{4\pi K\eta_1 } \nonumber\\
&\times&\int_{-\infty}^\infty \left[\cos\lambda\right]^{K-m} \left[\sin\lambda\right]^{m}
\frac{\sin(4K\eta_1\lambda)}{\lambda^2}d\lambda.\ \ \ 
\label{eq:betas}
\end{eqnarray}
\end{subequations}
In the calculations that leads to the set of equations
\eqref{eqs:MappingVoter} one assumes that the network elements $v_n$
are statistically independent and equivalent (see Appendix A). These
assumptions hold as long as the $K$ inputs of each element are chosen
randomly from anywhere in the system, namely, for the homogeneous
random topology. For other topologies that introduce correlations
between the network elements, such as the small-world or the
scale-free topologies, the mean-field assumptions do not necessarily
apply. However, when they do apply, the order parameter given in
Eq.~\eqref{eq:orderParameterVoter} becomes the sum of $N$ independent
and equally distributed random variables. Therefore, the determination
of $\psi(t)$ becomes analogous to determining the average position of
a 1D biased random walk, which can be solved exactly.

The stable fixed points of Eq.~\eqref{eq:mappingVoter1} give the
stationary values $\psi = \lim_{t\to\infty}\psi(t)$ of the order
parameter. It is clear from Eqs.~\eqref{eqs:MappingVoter} that $\psi =
0$ is always a fixed point. However, its stability depends on the
values of $\eta_1$ and $\eta_2$. Additionally, from
Eq.~\eqref{eq:betas} it follows that $\beta_m^K(\eta_1)=0$ for even
values of $m$ (because in such a case the integrand in that equation
is an odd function). Therefore, the polynomial in
Eq.~\eqref{eq:M-psi-Voter} contains only odd powers of $\psi(t)$ and
thus, for each fixed point $\psi$, the opposite value $-\psi$ is also
a fixed point.

To illustrate the formalism, we present here a detailed analysis of
the simple case $K=3$. However, the results are similar for any other
finite value of $K$. For $K=3$, the integrals in Eq.~\eqref{eq:betas}
can be easily computed (we used Mathematica \cite{wolfram}) and
Eq.~\eqref{eq:M-psi-Voter} becomes
\begin{subequations}\label{eqs:MK3}
\begin{equation}
M(\psi)=(1-2\eta_2)
\Big(\beta_1^3(\eta_1)\psi + \beta_3^3(\eta_1)\psi^3\Big),
\label{eq:M-psi-VoterK3}
\end{equation}
where
\begin{eqnarray}
\beta_1^3(\eta_1) &=& \frac{4+24\eta_1-|1-12\eta_1|-3|1-4\eta_1|}{32\eta_1},
\label{eq:beta31} \\
&&\nonumber\\
\beta_3^3(\eta_1) &=& -\frac{8\eta_1  - |1 - 12\eta_1| + |1 - 4 \eta_1|}{32\eta_1}.
\label{eq:beta33}
\end{eqnarray}
\end{subequations}
To determine the stability of the fixed points of the mapping
$M(\psi)$ we have to analyze the value of the derivative
$M'(\psi)\equiv dM(\psi)/d\psi$. If $|M'(\psi)|<1$ at the fixed point,
then that fixed point is stable. Otherwise, it is unstable. We further
divide our presentation in three cases.

\subsubsection{Sub-case 1: $\eta_2 = 0$}
Let us first show that the phase transition is discontinuous for the
case in which $\eta_2=0$, namely, when there is no intrinsic noise and
only the extrinsic noise is present. Under these circumstances, the
fixed-point equation $\psi=M(\psi)$ becomes
\begin{equation}
\psi = M(\psi) = \beta_1^3(\eta_1)\psi + \beta_3^3(\eta_1)\psi^3.
\label{eq:fixedPoint-K3}
\end{equation}
Using Eqs.~\eqref{eq:beta31} and \eqref{eq:beta33}, it is easy to see
that $\psi = 1$ and $\psi=-1$ are solutions of the fixed-point
equation \eqref{eq:fixedPoint-K3} provided that
$0\leq\eta_1<\frac{1}{4}$ (in addition to the trivial solution
$\psi=0$ which is always a fixed point). From Eqs.~\eqref{eq:beta31},
\eqref{eq:beta33} and \eqref{eq:fixedPoint-K3} we obtain that
\begin{eqnarray*}
M'(1) &=& \beta_1^3(\eta) + 3\beta_3^3(\eta_1)\\
& & \\
&=& \left\{
\begin{array}{lll}
0 & \mbox{ if } & 0\leq\eta_1\leq\frac{1}{12}\\
&& \\
\frac{12\eta_1 -1}{8\eta_1} & \mbox{ if } & \frac{1}{12}\leq\eta_1\leq\frac{1}{4}\\
&& \\
\frac{1}{4\eta_1} & \mbox{ if } & \frac{1}{4} \leq \eta_1 
\end{array}
\right.
\end{eqnarray*}

It follows from the above expression that $|M'(\pm1)|<1$ in the region
$0\leq\eta_1<\frac{1}{4}$, which shows that the fixed points $\psi=1$
and $\psi=-1$ are stable in this region. For $\eta_1>\frac{1}{4}$ the
fixed points $\psi=\pm1$ disappear and the only fixed point that
remains is $\psi=0$.

Let us compute now the stability of the fixed point $\psi=0$. From
Eqs.~\eqref{eq:beta31} and \eqref{eq:fixedPoint-K3} we get
\begin{eqnarray*}
M'(0) &=& \beta_1^3(\eta) \\
&& \\
&=& \left\{
\begin{array}{lll}
\frac{3}{2} & \mbox{ if } & 0\leq\eta_1\leq\frac{1}{12}\\
&& \\
\frac{12\eta_1 +1}{16\eta_1} & \mbox{ if } & \frac{1}{12}\leq\eta_1\leq\frac{1}{4}\\
&& \\
\frac{1}{4\eta_1} & \mbox{ if } & \frac{1}{4} \leq \eta_1 
\end{array}
\right.
\end{eqnarray*}
from which it follows that $|M'(0)| < 1$ for $\eta_1>\frac{1}{4}$,
whereas $|M'(0)| \geq 1$ for $0\leq\eta_1\leq\frac{1}{4}$.

The stability analysis presented above reveals that, when $\eta_2=0$,
the stable fixed points discontinuously transit from $\psi=\pm1$ to
$\psi=0$ as $\eta_1$ crosses the critical value $\eta_1^c=1/4$ from
below. Therefore, the phase transition in this case is discontinuous,
as is shown in Fig.~\ref{fig:phaseVoterK3}(b).

\subsubsection{Sub-case 2: $\eta_1 = 0$}

We consider now the case in which $\eta_1=0$, that is, when only
intrinsic noise is present. Taking the limit $\eta_1\to0$ in
Eqs.~\eqref{eq:beta31} and \eqref{eq:beta33} one gets
$\beta_1^3(0)=3/2$ and $\beta_3^3(0)=-1/2$. Therefore, in this case
Eq.~\eqref{eq:M-psi-VoterK3} becomes
\begin{equation}
M(\psi) = (1-2\eta_2)\left(\frac{3}{2}\psi-\frac{1}{2}\psi^3\right).
\label{eq:MK3-eta2-0}
\end{equation}
Let us start by analyzing the stability of the trivial fixed point
$\psi=0$. From the above equation we get
\[
 M'(0) = (1-2\eta_2)\frac{3}{2},
\]
from which it follows that $|M'(0)|<1$ only for $\eta_2 >
\frac{1}{6}$. Therefore, the disordered state characterized by the
fixed point $\psi=0$ is stable only for $\eta_2 > \frac{1}{6}$. As
$\eta_2$ decreases below the critical value $\eta_2^c=\frac{1}{6}$,
the disordered phase becomes unstable and two stable non-zero fixed
points appear. Assuming $\psi\neq0$, the fixed point equation
$\psi=M(\psi)$ can be solved for $\psi$ obtaining
\[
\psi = \pm\left(3-\frac{2}{1-2\eta_2}\right)^{1/2}.
\]

A stability analysis reveals that the above fixed points are stable
for $\eta_2<\eta_2^c$ (in this region $|M(\psi)|<1$) and unstable for
$\eta_2>\eta_2^c$ (because in this other region
$|M(\psi)|>1$). Summarizing, the stable fixed points for the case
$\eta_1=0$ are
\[
\psi = \left\{
\begin{array}{lll}
\pm\left(3-\frac{2}{1-2\eta_2}\right)^{1/2} & \mbox{ if } & 0\leq\eta_2\leq\eta_2^c\\
& & \\
0 & \mbox{ if } & \eta_2^c< \eta_2
\end{array}
\right.
\]
where $\eta_2^c = 1/6$. This result is plotted in
Fig.~\ref{fig:phaseVoterK3}(a) (solid line), from which it is apparent
that the phase transition in the majority voter model with only
intrinsic noise is indeed continuous. Additionally, for values of
$\eta_2$ below, but close to, the critical value $\eta_2^c$ at which
the phase transition occurs, the order parameter $\psi$ behaves as
$\psi\approx\pm(\eta_2^c -\eta)^{1/2}$, which shows that this phase
transition belongs to the mean-field universality class.

\begin{figure}[t]
\scalebox{0.95}{\includegraphics{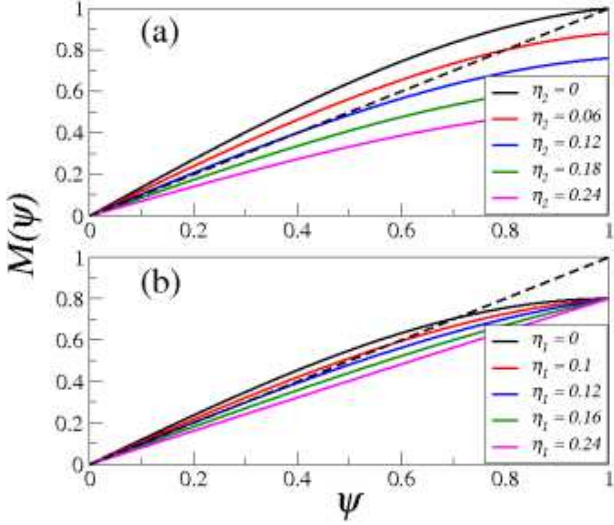}}
\caption{(Color online) Graph of the dynamical mapping $M(\psi)$ as a
function of $\psi$ for $K=3$. (a) $\eta_1=0.1$ and different values of
$\eta_2$; (b) $\eta_2=0.1$ and different values of $\eta_1$. Note that
in both cases, as the noise intensity decreases the stable nonzero
fixed point appears continuously. This is always the behavior for
finite values of $K$.}
\label{fig:MVoterK3}
\end{figure}

\begin{figure}[t]
\scalebox{0.95}{\includegraphics{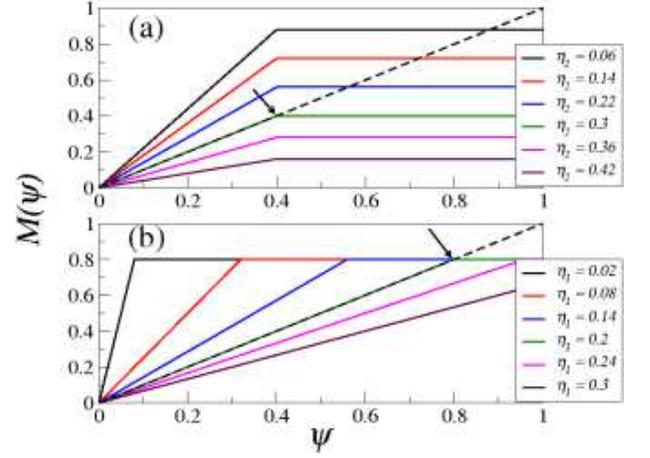}}
\caption{(Color online) Graph of the dynamical mapping $M(\psi)$ as a
function of $\psi$ for $K\to\infty$. (a) $\eta_1=0.1$ and different
values of $\eta_2$; (b) $\eta_2=0.1$ and different values of
$\eta_1$. Note that in both cases, as the noise intensity decreases
the stable nonzero fixed point appears discontinuously (indicated by
the arrows). For infinite $K$, the phase transition is always
discontinuous.}
\label{fig:MVoterKinf}
\end{figure}

\subsubsection{Sub-case 3: $\eta_1\neq0$ and $\eta_2\neq0$}

When both types of noise, intrinsic and extrinsic, are present in the
system, the phase transition is always continuous for any finite value
of $K$. To illustrate this we present in Fig.~\ref{fig:MVoterK3}(a)
the graph of $M(\psi)$ for $\eta_1=0.1$ and different values of
$\eta_2$. Note that $M(\psi)$ is a monotonically increasing convex
function, and therefore the nonzero stable fixed point appears
continuously as $\eta_2$ decreases. The same happens if we now fix the
value of $\eta_2$ and vary the value of $\eta_1$, as it is shown in
Fig.~\ref{fig:MVoterK3}(b). This behavior is typical of a second order
phase transition.

\begin{figure*}[ht]
\scalebox{0.95}{\includegraphics{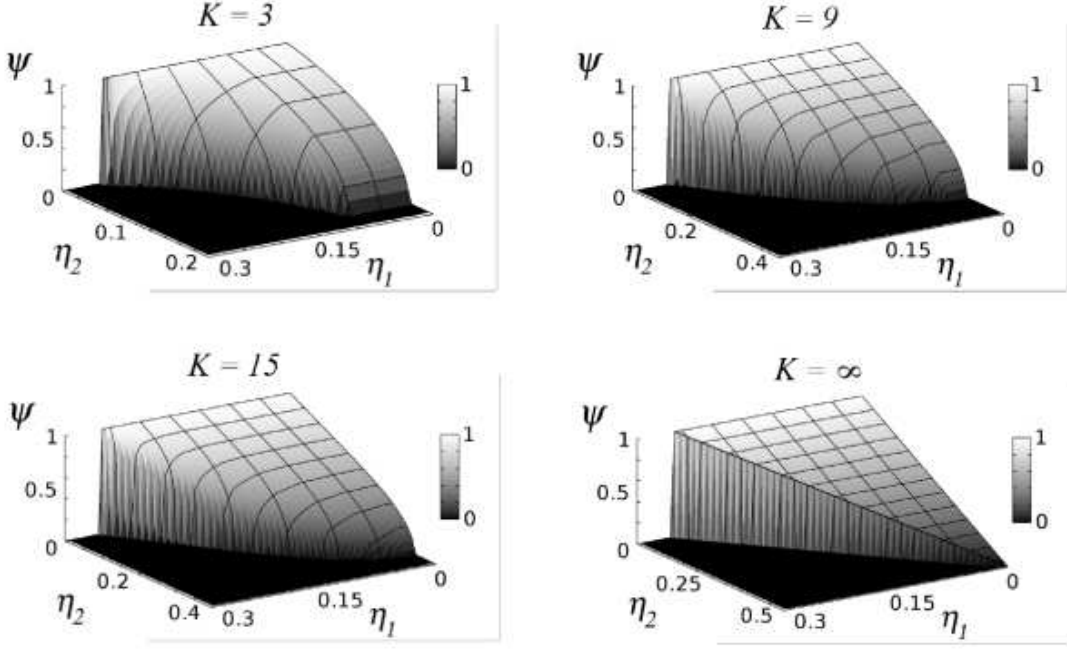}}
\caption{Phase transition in the majority voter model. Order parameter
$\psi$ as a function of the extrinsic noise $\eta_1$ and the intrinsic
noise $\eta_2$ for $K=3$, $K=9$, $K=15$ and $K\to\infty$. Note that for
any finite value of $K$ the phase transition is always continuous
except for the case $\eta_1=0$. Note also that the phase transition it
is always discontinuous for $K\to\infty$.}
\label{fig:PVoterKFinite}
\end{figure*}

Assuming $\psi\neq0$ and using Eq.~\eqref{eq:M-psi-VoterK3}, the fixed
point equation $\psi = M(\psi)$ can be solved for $\psi$ obtaining
\begin{equation}
\psi = \pm\left\{\frac{1}{\beta_3^3(\eta_1)}\left(\frac{1}{1-2\eta_2}
-\beta_1^3(\eta_1)\right)\right\}^{1/2}
\label{eq:psiStableVoterK3}
\end{equation}
For this equation to have real solutions the quantity inside the
curly brackets must be positive. From Eq.~\eqref{eq:beta33} it
follows that $\beta_3^3(\eta_1) \leq0$ for any positive value of
$\eta_1$. Therefore, Eq.~\eqref{eq:psiStableVoterK3} has real
solutions only if
\[
\frac{1}{1-2\eta_2}\leq\beta_1^3(\eta_1).
\]
The values of $\eta_1$ and $\eta_2$ for which the equality holds in
the above expression determine the critical line on the
$\eta_1$-$\eta_2$ plane at which the phase transition
occurs. Fig.~\ref{fig:PVoterKFinite} shows surface plots for the
(positive) value of the stable fixed point $\psi$ as a function of
$\eta_1$ and $\eta_2$ for $K=3$, $K=9$, $K=15$ and
$K\to\infty$. Interestingly, for any finite value of $K$ the phase
transition is \emph{always} continuous except for the special case
$\eta_2=0$. Therefore, for any finite $K$, even a small amount of
intrinsic noise suffices to make the phase transition continuous.

\subsection{Case 2: $K = \infty$}

In Appendix A we show that for $K\to\infty$ the temporal evolution of
the order parameter is still given by the dynamical mapping
Eq.~\eqref{eq:mappingVoter1}, where now $M(\psi)$ is
\begin{equation}
M(\psi) =
\left\{ 
\begin{array}{rlr}
-(1-2\eta_2) & \mbox{ if } & \psi < -4\eta_1\\
&&\\
\frac{1-2\eta_2}{4\eta_1}\psi & \mbox{ if } & |\psi|\leq 4\eta_1\\
&&\\
1-2\eta_2 & \mbox{ if } & \psi > 4\eta_1
\end{array}
\right.
\label{eq:discontVoter}
\end{equation}
Fig.~\ref{fig:MVoterKinf}(a) shows the behavior of $M(\psi)$ for
$K\to\infty$, $\eta_1=0.1$ and different values of $\eta_2$, and
Fig.~\ref{fig:MVoterKinf}(b) shows the same kind of plots but now
keeping $\eta_2=0.1$ and varying the value of $\eta_1$. It can be seen
from Fig.~\ref{fig:MVoterKinf}(a) that the phase transition is
discontinuous. Indeed, as $\eta_2$ decreases below the critical value
$\eta_2^c=0.3$, the non-zero stable fixed point appears
discontinuously (see the point indicated with an arrow in the figure).
An analogous behavior occurs in Fig.~\ref{fig:MVoterKinf}(b) when
$\eta_1$ reaches the value $\eta_1^c=0.2$. Thus, in the limit
$K\to\infty$ the phase transition is \emph{always} discontinuous (see
Fig.~\ref{fig:PVoterKFinite}). In this sense, the discontinuity in the
phase transition observed when only extrinsic noise is used can be
considered as a singular limit, either $\eta_2\to0$ or $K\to\infty$,
of a phase transition that is otherwise continuous.

\section{The vectorial network model}
\label{sec:vectorial}

The second network model that we analyze, which we will call the
\emph{Vectorial Network Model} (VNM), is much closer to the
self-propelled model than the voter model presented in the
previous section. As we will see later, the VNM corresponds to a
mean-field theory of the self-propelled model. It consists of a
network with $N$ nodes (or elements) which, as in the self-propelled
model, are the two dimensional vectors
$\vec{v}_1=e^{i\theta_1},\dots,\vec{v}_N=e^{i\theta_N}$ (represented
as complex numbers). All the vectors have the same magnitude
$|\vec{v}_n|=1$ but their orientations $\theta_1,\dots\theta_N$ in the
plane can change. Each vector $\vec{v}_n$ is connected to a fixed set
of $k_n$ other vectors, ${\mathcal I}_n=\{\vec{v}_{n_1},
\vec{v}_{n_2}, \dots, \vec{v}_{n_{k_n}}\}$, from which $\vec{v}_n$
will receive an input signal. We will call this set \emph{the inputs}
of $\vec{v}_n$, and consider again the homogeneous random topology in
which all the elements have exactly $K$ inputs chosen randomly from
anywhere in the system. The input signal $\vec{U}_n(t)$ received by
$\vec{v}_n$ from its $K$ inputs is defined as
\begin{equation}
\vec{U}_n(t) = \frac{1}{K}\sum_{j=1}^K \vec{v}_{n_j}(t)
\label{eq:inputSignalVec} 
\end{equation}

For the interaction between the network elements we consider from the
beginning a dynamic rule that already incorporates both types of
noise, intrinsic and extrinsic:
\begin{equation}
 \theta_n(t+1) = \mbox{Angle}\left[\vec{U}_n(t) + \eta_1 e^{i\xi_n(t)}\right]+\eta_2\zeta_n(t), 
\label{eq:voterRule} 
\end{equation}
where $\xi_n(t)$ and $\zeta_n(t)$ are independent random variables
uniformly distributed in the interval $[-\pi,\pi]$. The noise
intensities $\eta_1$ and $\eta_2$, which take constant values between
0 and 1, are the amplitudes of the extrinsic (Gr\'egoire-Chat\'e) and
intrinsic (Vicsek) types of noise, respectively. Note that, while in
the self-propelled model the particles can move and thus the vectors
$\vec{v}_n$ represent particle velocities, in the VNM the particles do
not move. Rather, they are fixed to the nodes of the network. For this
reason, in the VNM the vectors $\vec{v}_n$ cannot be considered as
velocities, but just as a given property of the particles (such as
spin).

\subsection{Equivalence between the VNM and the self-propelled model}

The main difference between the self-propelled model and the VNM is
that in the former the motion of the particles can produce
correlations in space and time that couple the global order with the
local density, affecting the phase transition. Such coupling is not
present in the VNM unless we choose very specific network topologies
that change over time. However, here we are interested only on the
effects that the two different types of noise have on the phase
transition. The VNM is especially suited for this analysis precisely
because of the absence of such complicated dynamical effects as the
coupling between order and density.

\begin{figure}[t]
\scalebox{0.95}{\includegraphics{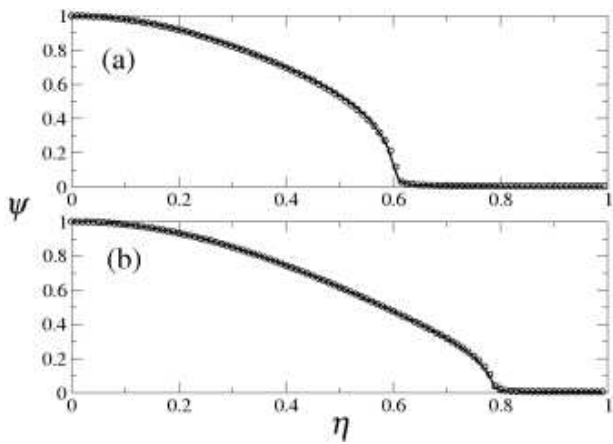}}
\caption{Phase transition for the VNM with intrinsic noise only (solid
line) and the SPMIN with random mixing (symbols), both with
$N=20000$. For the SPMIN the size of the box is $L=32$ and the radius $r$ of the interaction
vicinity has been chosen so that the average number of
interactions per particle coincides with the network connectivity $K$
in the VNM. The two panels correspond to (a) $K=5$ ($r\approx0.285$)and (b)
$K=20$ ($r\approx0.571$). Note that the phase transition in this case is continuous and
the same for both the VNM and the SPMIN with random
mixing.}
\label{fig:randomMixing1}
\end{figure}

Nonetheless, it is important to note that the limit of large particle
speeds of the self-propelled model is well described by the
VNM. Indeed, if the speed of the particles in the self-propelled model
is small, then particles that were within the same interaction
vicinity at time $t$ will most likely remain within the same
interaction vicinity at the next time step $t+\Delta t$. Therefore,
for small particle speeds the spatial correlations between the
particles are important. Contrary to this, in the opposite limit of
large particle speeds, and because of the noise in the direction of
motion of each particle (whether intrinsic or extrinsic), particles
that at time $t$ were within the same vicinity will most likely not
remain within that vicinity and end up interacting with different
particles at the next time step $t+\Delta t$. Therefore, in the limit
$v\to\infty$ of the self-propelled model spatial correlations are
lost in only one time step. This is equivalent to randomly mixing
the particles within the box at every time step, which is precisely
the condition for the mean-field theory conveyed in the VNM to be
exactly applicable.

\begin{figure}[t]
\scalebox{0.95}{\includegraphics{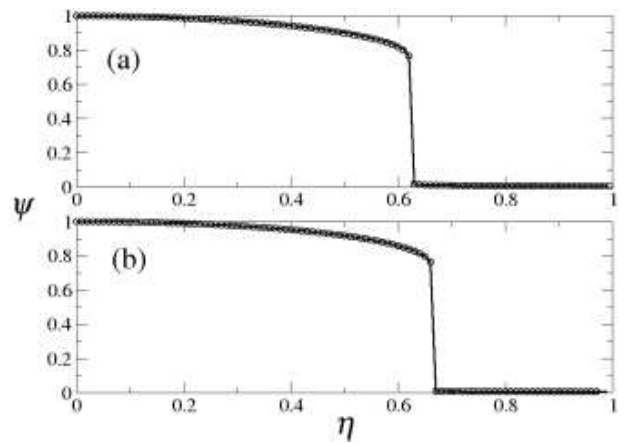}}
\caption{Phase transition for the VNM with extrinsic noise only (solid
line) and the SPMEN with random mixing (symbols). The systems have the same 
parameters as in Fig.~\ref{fig:randomMixing1}: $N=2\times10^4$, $L=32$, $K=4$ ($r\approx0.285$) in (a)
and $K=20$ ($r\approx0.571$) in (b). Note that the phase transition in this case is discontinuous
and the same for both the VNM and the SPMEN with random mixing.}
\label{fig:randomMixing2}
\end{figure}

Fig.~\ref{fig:randomMixing1} shows the value of the order parameter
$\psi$ as a function of the noise intensity $\eta$ for the VNM with
intrinsic noise only (solid line), and for the SPMIN \emph{with random
mixing} (symbols) \footnote{The order parameter $\psi(t)$ for the VNM
is defined exactly in the same way as for the self-propelled model
(see Eq.~\eqref{eq:orderParameter}).}. Namely, instead of updating the
positions of the particles in the SPMIN according to the kinematic
rule given in Eq.~\eqref{eq:kinetic1}, we just randomly mixed all the
particles within the box at every time step. We used equivalent
systems with $N=20000$ particles and adjusted the other parameters of
the SPMIN in such a way that the average number of interactions per
particle $K$ was the same as for the vectorial network model
[$K=5$ in Fig.~\ref{fig:randomMixing1}(a) and $K=20$ in
Fig.~\ref{fig:randomMixing1}(b)]). Analogously,
Fig.~\ref{fig:randomMixing2} shows equivalent results but for the VNM
with extrinsic noise only (solid line) and the SPMEN with random mixing (symbols). As we
can see from these figures, regardless of the type of noise both the
self-propelled model with random mixing and the VNM give the same
phase transition within numerical accuracy.

Although the above results do not constitute a proof, they do suggest
that, in the limit of large particle speeds, the self-propelled model
becomes equivalent to the VNM. It has been claimed that this
equivalence is obtained only through a \emph{singular limit}
\cite{ChateComment}. Though this might be the case, in our numerical
simulations with different system sizes we always obtain results
analogous to the ones displayed in Figs.~\ref{fig:randomMixing1} and
\ref{fig:randomMixing2}, which do not seem to show such singular
behavior.

\subsection{Mean-field theory of the VNM}

Let us define the average vector $\vec{U}(t)$ as
\begin{equation}
\vec{U}(t) = \frac{1}{N}\sum_{n=1}^N \vec{v}_n(t).
\label{eq:defU}
\end{equation}
In the context of the self-propelled model, $\vec{U}(t)$
is the average velocity of the entire system. Clearly, the
instantaneous value of the order parameter $\psi(t)$ is related to
this vector through $\psi(t) =
\left|\vec{U}(t)\right|$. Let $(\psi(t),\theta(t))$ be
the polar coordinates of $\vec{U}(t)$, and
$P_{\vec{U}}(\psi,\theta;t)$ its probability distribution function (in
polar coordinates). Note, then, that $\psi(t)$ is the first radial
moment of $P_{\vec{U}}(\psi,\theta;t)$.

\begin{figure*}[t]
\scalebox{0.95}{\includegraphics{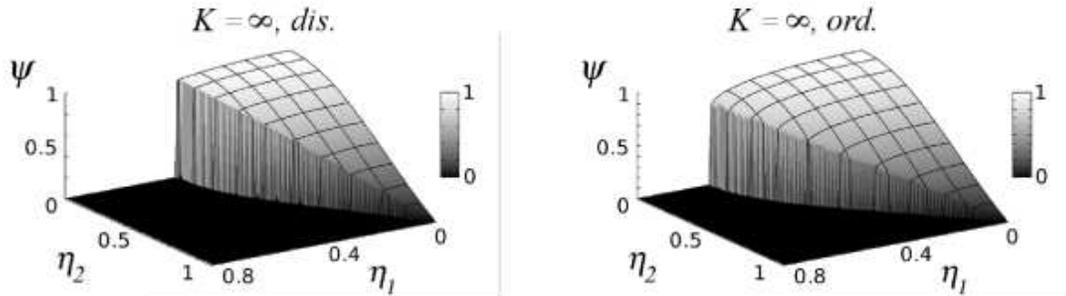}}
\caption{Phase transition in the VNM with $K\to\infty$. The left panel
correspond to the phase transition obtained starting out the dynamics
from a fully disordered condition ($\psi(0)\approx0$), whereas for the
panel on the right the dynamics were started from fully ordered
initial conditions ($\psi(0)\approx1$). In both cases the phase
transition is discontinuous for any non-zero value of the extrinsic
noise amplitude $\eta_1$. These plots were obtained by numerically
finding the fixed points of $M(\psi)$ given in
Eq.~\eqref{eqs:mapVecKinfty}}.
\label{fig:psiVecKInfty}
\end{figure*}

As in the voter model, we assume that the network elements
$\vec{v}_n$ are statistically independent and equivalent. Therefore,
Eq.\eqref{eq:defU} becomes the sum of $N$ independent and equally
distributed random variables, and the problem of determining
$\vec{U}(t)$ is then similar to that of finding the
position of a 2D biased random walk.  In Appendix~B we present this
mean-field calculation, and show that the temporal evolution of the
Fourier transform of $P_{\vec{U}}(\psi,\theta;t)$ is given by the
recurrence relation
\begin{eqnarray}
 && \hat{P}_{\vec{U}}(\lambda,\gamma;t+1) = J_0(\lambda) +
 \sum_{m=-\infty}^\infty \frac{\sin(m\pi\eta_2)}{2\pi^2\eta_2}
 J_m(\lambda)e^{im\gamma} \nonumber\\ &\times& \int_{0}^\infty
 \frac{d\lambda'}{\lambda'}J_0(K\eta_1\lambda')\int_{0}^{2\pi}d\gamma'
 \left[\hat{P}_{\vec{U}}(\lambda',\gamma';t)\right]^K e^{-im\gamma'},
\label{eq:recurMainText}
\end{eqnarray}
where the $J_m(x)$'s are Bessel functions, and $\lambda$ and $\gamma$
are the Fourier conjugate variables to $\psi$ and $\theta$
respectively. In principle, this equation can be solved for any finite
value of the network connectivity $K$. However, we were able to solve
it only in the limit case $K\to\infty$. Therefore, we present first
the analytic results for $K\to\infty$. In the next section we present
numerical results for finite values of $K$.

\subsection{Case 1: $K\to\infty$}

In the limit $K\to\infty$ the factor
$\left[\hat{P}_{\vec{U}}(\lambda',\gamma';t)\right]^K$ appearing in
the second integral of Eq.~\eqref{eq:recurMainText} can be replaced by
a Dirac delta function radially centered at $K\psi(t)$. This leads to
\begin{eqnarray}
 \hat{P}_{\vec{U}}(\lambda,\gamma;t+1) &=& J_0(\lambda) +
 \sum_{m=-\infty}^\infty (-i)^m \frac{\sin(m\pi \eta_2)}{\pi\eta_2}
 J_m(\lambda) \nonumber\\ &\times& e^{im(\gamma-\alpha)}
 \int_{0}^\infty \frac{dx}{x}J_0(\eta x) J_m\left(\psi(t) x\right),
\label{eq:recur2MainText}
\end{eqnarray}
From this equation it follows that the order parameter $\psi(t)$,
which is the first radial moment of $P_{\vec{U}}(\psi,\theta;t)$,
obeys the dynamical mapping (see details in Appendix B)
\begin{equation}
 \psi(t+1) = \frac{\sin(\pi\eta_2)}{\pi\eta_2}\int_0^\infty J_0(\eta_1 x) J_1\left(\psi(t) x\right)\frac{dx}{x}.
\label{eq:recur3MainText}
\end{equation}

The integral on the right-hand side of the above equation is an
instance of the Weber-Schafheitlin integrals
\cite{A_and_S}. After the evaluation of this integral, the dynamical mapping
for the order parameter can be written as
\begin{subequations}\label{eqs:mapVecKinfty}
\begin{equation}
 \psi(t+1) = M\left(\psi(t)\right),
\label{eq:MVectorial1}
\end{equation}
where the mapping $M(\psi)$ is 
\begin{equation}
 M(\psi) = \left\{
\begin{array}{lll}
 \frac{\sin(\pi\eta_2)\psi(t)}{2\pi\eta_1\eta_2}
{}_2F_1\left(\frac{1}{2},\frac{1}{2};2,\left[\frac{\psi}{\eta_1}\right]^2\right)
& \mbox{ if } & \psi < \eta_1 \\ & & \\
\frac{\sin(\pi\eta_2)}{\pi\eta_2}{}_2F_1\left(\frac{1}{2},-\frac{1}{2};1,\left[\frac{\eta_1}{\psi}\right]^2
\right)& \mbox{ if } & \psi > \eta_1
\end{array}\right.
\label{eq:MVectorial2}
\end{equation}
\end{subequations}
and ${}_2F_1(a,b;c,d)$ are hypergeometric functions. The fixed points
of $M(\psi)$ give the stationary value $\psi$ of the order
parameter. In Ref.~\cite{OurPRL} we have shown that for $\eta_2=0$,
the non-trivial fixed point of this mapping appears discontinuously as
$\eta_1$ crosses the critical value $\eta_1\approx0.672$ from
above. On the other hand, for $\eta_2>0$ there is a global factor
$\frac{\sin(\pi\eta_2)}{\pi\eta_2}$ which does not change the
discontinuous appearance of the non-trivial fixed point. Therefore,
for any values of $\eta_1$ and $\eta_2$, the phase transition is
discontinuous.

\begin{figure}[t]
\scalebox{0.95}{\includegraphics{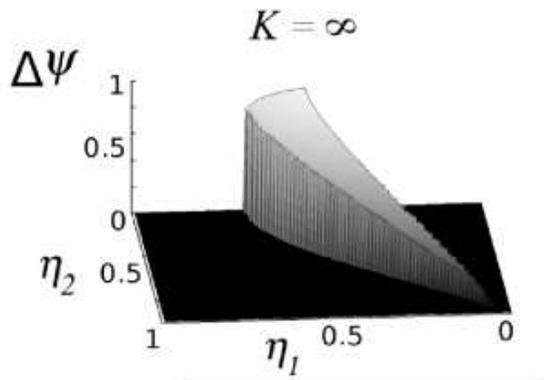}}
\caption{Difference $\Delta \psi = \psi_{ord}-\psi_{dis}$ between the
two surface plots displayed on Fig.~\ref{fig:psiVecKInfty}. The region
for which $\Delta\psi\neq0$ is the region of metastability where the
system exhibits hysteresis. Note that this region crosses the entire
square $[0,1]\times[0,1]$ in the $\eta_1$-$\eta_2$ plane from one side
to the other.}
\label{fig:diffPsiVecKInfty}
\end{figure}

Fig.~\ref{fig:psiVecKInfty} shows surface plots of the stationary
value $\psi$ as a function of $\eta_1$ and $\eta_2$ for two different
cases: (i) when the dynamics of the VNM start out from disordered
initial conditions [$\psi(0)\approx0$ in Eq.~\eqref{eq:MVectorial1}],
and (ii) when the dynamics start out from ordered initial conditions
[$\psi(0)\approx1$ in Eq.~\eqref{eq:MVectorial1}]. Let us denote as
$\psi_{dis}$ and $\psi_{ord}$ the stationary values of the order
parameter obtained in each of the two cases mentioned above,
respectively. It is apparent from Fig.~\ref{fig:psiVecKInfty} that
$\psi_{ord}$ and $\psi_{dis}$ are equal in a large region of the
$\eta_1$-$\eta_2$ parameter space.  However, there is also a region in
which $\psi_{ord}$ and $\psi_{dis}$ are different. This latter region,
where the system shows hysteresis, is shown in
Fig.~\ref{fig:diffPsiVecKInfty}, in which the difference $\Delta\psi =
\psi_{ord}-\psi_{dis}$ is plotted as a function of $\eta_1$ and
$\eta_2$. The region for which $\Delta\psi\neq0$ is a region of
metastability where two stable fixed points exist, the trivial one
$\psi_{dis}=0$ and the non-zero fixed point $\psi_{ord}$. It is clear
from these results that the VNM with $K\to\infty$ exhibits a
discontinuous phase transition for any non-zero value of the extrinsic
noise $\eta_1$. On the other hand, for $\eta_1=0$, the amount of order
in the system decreases as $\eta_2$ increases. However, there is no
phase transition in this case since $\psi=0$ only when $\eta_2$
reaches its maximum value $\eta_2=1$. (In ferromagnetic systems,
$\eta_2=1$ would correspond to infinite temperature.) In other words,
for infinite connectivity and zero extrinsic noise, the order in the
system can never be destroyed by the intrinsic noise, unless it
reaches its maximum value. However, in the presence of both types of
noise, extrinsic and intrinsic, the phase transition is always
discontinuous.

\subsection{Case 2: $K$ finite}

As it was mentioned before, we do not have an analytic solution of
Eq.~\eqref{eq:recurMainText} for finite values of $K$. Nonetheless,
numerical simulations show a phase transition that is continuous in one
region of the $\eta_1$-$\eta_2$ parameter space, and
discontinuous in another region. This is qualitatively different from
the phase transition observed in the majority voter model, which was
always continuous for any finite value of $K$ (except for $\eta_2=0$).

\begin{figure*}[t]
\scalebox{0.95}{\includegraphics{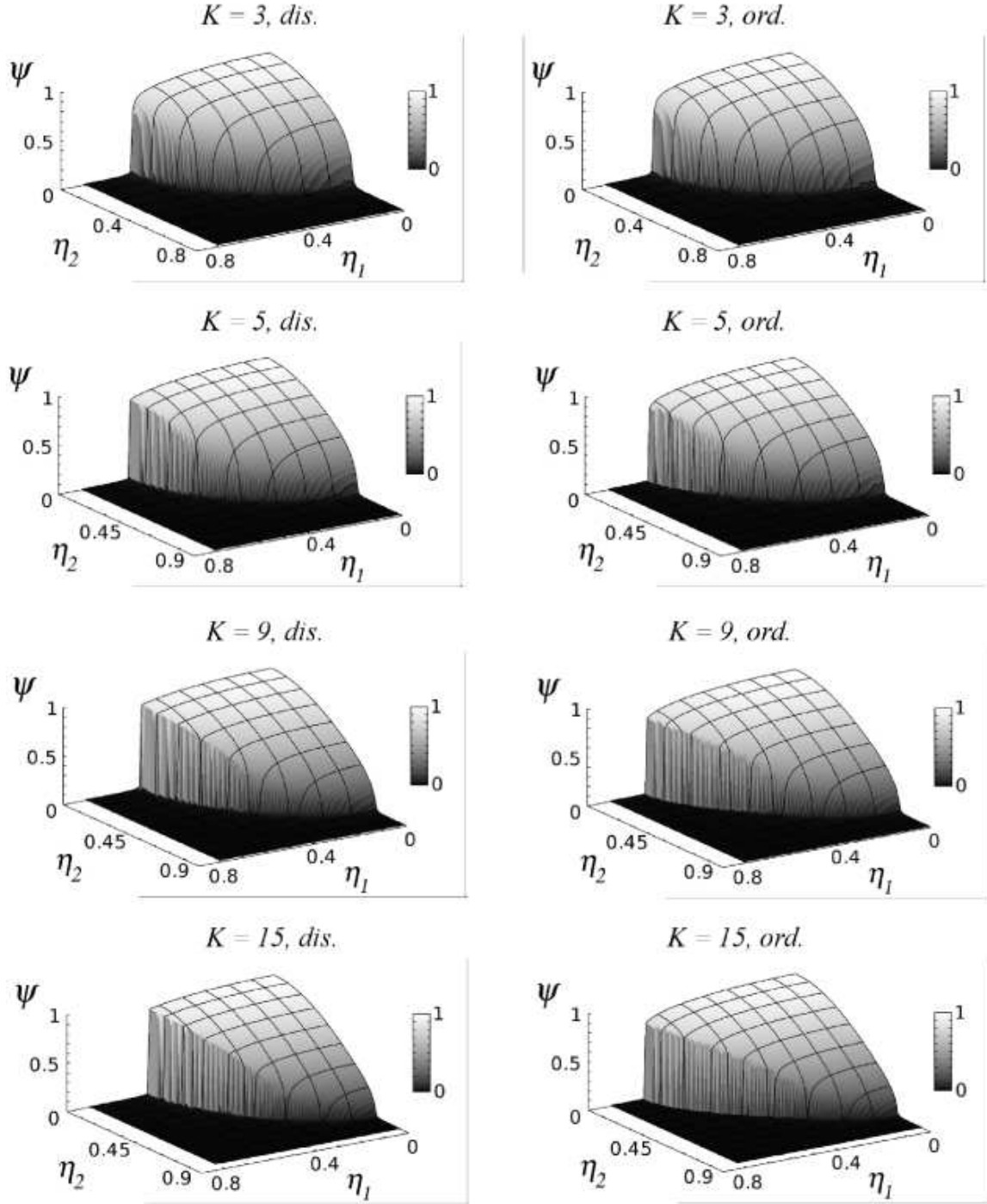}}
\caption{Surface plots show the order parameter $\psi$ as a function
of the extrinsic and extrinsic noise amplitudes $\eta_1$ and $\eta_2$
respectively, for $K=3$, $K=5$, $K=9$ and $K=15$.  The graphs were
obtained through numerical simulation for systems with $K=2000$
elements. The panels on the left correspond to random initial
conditions, whereas the panels on the right correspond to fully
ordered initial conditions. Note that, except for the case $K=3$,
there is a region of hysteresis where the phase transition is
discontinuous. }
\label{fig:psiVecKFinite}
\end{figure*}

\begin{figure*}[t]
\scalebox{0.95}{\includegraphics{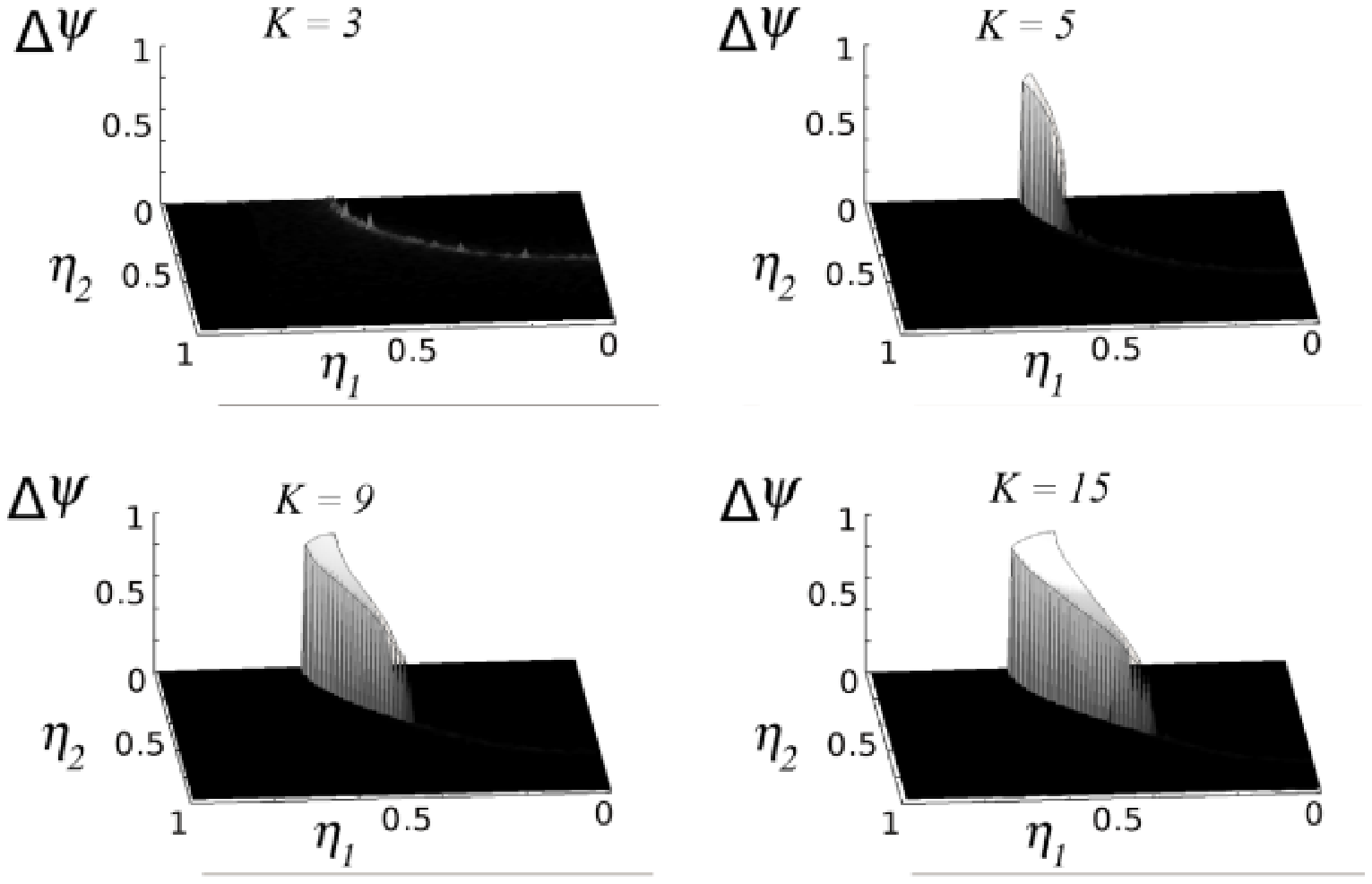}}
\caption{Difference $\Delta \psi = \psi_{ord}-\psi_{dis}$ between the
surface plots displayed on Fig.~\ref{fig:psiVecKFinite}. The region of
metastability where $\Delta\psi\neq0$ grows with $K$. However, for any
finite value of $K$ this region does not cross the entire square
$[0,1]\times[0,1]$ in the $\eta_1$-$\eta_2$ plane.}
\label{fig:diffPsiVecKFinite}
\end{figure*}

Fig.~\ref{fig:psiVecKFinite} shows $\psi$ as a function of $\eta_1$
and $\eta_2$ for different values of $K$. The results reported in this
figure were obtained through numerical simulations of the VNM for
systems with $N=20000$. The figures on the left correspond to
disordered initial conditions ($\psi(0)\approx0$), whereas those on
the right correspond to ordered initial conditions
($\psi(0)\approx1$). The difference $\Delta\psi=\psi_{ord}-\psi_{dis}$
as a function of $\eta_1$ and $\eta_2$ is plotted on
Fig.~\ref{fig:diffPsiVecKFinite}. It is apparent from these figures
that, except for the case $K=3$, there is a region of hysteresis where
$\Delta\psi\neq0$. This region grows as $K$ increases, but it does not
seem to cross the square $[0,1]\times[0,1]$ of the $\eta_1$-$\eta_2$
parameter space from one side to the other (as it does for
$K\to\infty$). This implies that the phase transition is discontinuous
along the boundary of the region in which $\Delta\psi\neq0$, but it is
continuous along the boundary where $\psi\to0^+$ and
$\Delta\psi=0$. 

It is worth emphasizing that for $\eta_1=0$, namely when there is no
extrinsic noise, the VNM always undergoes a continuous phase transition from
ordered to disordered states as the intensity of the intrinsic noise
$\eta_2$ increases. This is consistent with the behavior originally
reported by Vicsek et. al. for the SPMIN [see
Fig.~\ref{fig:phaseVicsek}(a)]. On the other hand, for $\eta_2=0$,
i.e. in the absence of intrinsic noise, the phase transition in the
VNM is discontinuous as a function of the extrinsic noise amplitude
$\eta_1$, which is consistent with the phase transition observed in
the SPMEN [see Fig.~\ref{fig:phaseVicsek}(b)].

\section{What do we mean by ``mean-field'' theory?}
\label{sec:mean-field}

In the computations of these two network models we have used the
``mean field'' assumption that all the spins are statistically
independent and statistically equivalent, though we consider arbitrary
values of $K$ (the number of particles that interact). This leads to
phase transitions that can be continuous or discontinuous depending
upon the values of $\eta_1$ and $\eta_2$. However, a frequent (and
frequently equivalent) view of what constitutes a ``mean field''
theory entails the assumption that every particle interacts with all
the other particles in the system. This second assumption is akin to
the case $K\to\infty$, for which the phase transition is always
discontinuous and, in general, of a different nature than the one
obtained for finite $K$.

In our network models, the assumption of statistical independence is
certainly true for the homogeneous random topology in which the $K$
inputs of each element are randomly chosen from anywhere in the
system. In this case, the probability for two distinct elements
$\vec{v}_m$ and $\vec{v}_n$ to have at least one of its $K$ inputs in
common is of order $1/N$. In the thermodynamic limit $N\to\infty$
this probability is zero. Therefore, for large systems all the
elements have different sets of inputs and are indeed statistically
independent.

\begin{figure}[t]
\scalebox{0.95}{\includegraphics{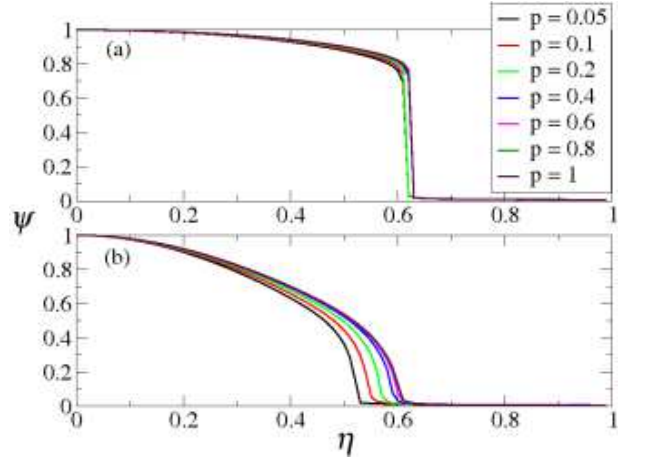}}
\caption{(Color online.) Phase transition in the VNM on small-world
networks. (a) In the absence of intrinsic noise, the extrinsic noise
produces a discontinuous phase transition regardless of the rewiring
connectivity $p$.  (b) The phase transition is continuous for all
values of $p$ when only the intrinsic noise is present.}
\label{fig:small-world}
\end{figure}

For other topologies, such as small-world, a large fraction of the
network elements share inputs even in the thermodynamic limit. For
this topology the assumption of statistical independence is no longer
valid.  However, the qualitative behavior of the phase transition
obtained for the VNM on small-world networks, in which the short-range
interactions induce spatial correlations between the network elements,
is similar to the one observed for the homogeneous random topology
described so far.  Fig.~\ref{fig:small-world} shows the phase
transition in the VNM on small-world networks. To generate this figure
we used the standard Watts-Strogatz small-world algorithm
\cite{Watts3,Watts4,Strogatz}, placing the elements on a 2D square
lattice. Initially each element $\vec{v}_n$ has $K=5$ inputs, which
are chosen as its four first-neighbors and the element $\vec{v}_n$
itself (i.e., we allow self-interactions). Then, with probability $p$
each input connection in the network is rewired to a randomly chosen
element. Thus, if $p=0$ we have a regular square lattice, whereas if
$p=1$ the topology becomes homogeneously random. This latter case is
where our mean-field theory results are exactly applicable.

Fig.~\ref{fig:small-world}(a) corresponds to the case in which there
is no intrinsic noise ($\eta_2=0$) and only the extrinsic noise is
present, and Fig.~\ref{fig:small-world}(b) shows the opposite case
where there is no extrinsic noise ($\eta_1=0$). The different curves
in each figure correspond to different values of the rewiring
probability $p$. Note that the nature of phase transition,
i.e. whether continuous or discontinuous, does not change with the
rewiring probability $p$. What changes is the critical value of the
noise at which the phase transition occurs, but the continuity of the
phase transition does not change with $p$. This is important because
for small values of $p$ there are strong spatial correlations in the
system generated by the first-neighbor interactions in the small-world
network. However, even in the presence of such spatial correlations,
the phase transition appears to be discontinuous for the extrinsic
noise and continuous for the intrinsic noise. One advantage of the 
VNM is that the finite-size effects are much smaller than in the
self-propelled model. Therefore, the continuous or discontinuous
character of the phase transiion can be very well observed with
$N=20000$ particles. 

\section{Summary and discussion}
\label{sec:conclusions}

Although the self-propelled model proposed by Vicsek et. al. is one of
the simplest models that we have to describe the emergence of
collective order in groups of organisms, its analytic solution has
remained elusive for more than a decade. Due to this lack of an
analytical formalism to analyze the dynamical properties of the
self-propelled model, it has not been possible to properly
characterize some of its most basic features. In this work we have
been particularly interested in how the way in which the noise is
introduced in the system affects the nature of the phase
transition. In the self-propelled model, numerical simulations
indicate that the intrinsic noise originally proposed by Vicsek and
his group seems to produce a continuous phase transition. In contrast,
the extrinsic noise later introduced by Gr\'egoire and Chat\'e
generates a phase transition that is clearly discontinuous.

To determine the effect that each type of noise has on the phase
transition, we have presented two network models that capture the main
characteristics of the interactions in the self-propelled model and
that can be handled analytically. These two network models, which we
call the majority voter model and the vectorial network model, can be
considered as the mean-field representation of the self-propelled
model in 1 and 2 dimensions. In fact, we have shown numerically that
the self-propelled model with random mixing, which is obtained in the
limit of large particle speeds, is fully equivalent to the vectorial
network model.

When the number of interactions per particle is finite, our numeric
and analytic results for the two network models show that, in the
absence of extrinsic noise, the phase transition driven by the
intrinsic noise is continuous, in accordance with the results obtained
for the original Vicsek model (SPMIN). In the opposite case, namely
when there is no intrinsic noise and only the extrinsic noise is
present, the phase transition is discontinuous, which is consistent
with the behavior reported by Gr\'egoire and Chat\'e for the
SPMEN. The above results are true even for the small-world topology in
which there exist strong spatial correlations induced by the
first-neighbor connections.

For intermediate cases in which both types of noise are present, and
for finite network connectivities, the situation is more complicated:
The voter model exhibits a phase transition that is always continuous,
whereas for the vectorial model there is a region in the
noise-parameter space where the phase transition is continuous, and
another region where it is discontinuous. In this later case, the
region of discontinuity grows as the network connectivity
increases. Interestingly, in the limit of infinite network
connectivities and in the presence of both types of noise, the phase
transition is always discontinuous in both the voter model and the
vectorial model.

An interesting open question that follows from our work is to what
extent can the stationary collective features of the network systems
agree quantitatively with those of the corresponding swarming
models. It has been suggested that the coupling between global order
and local density could make the phase transition in the
self-propelled model discontinuous regardless of the type of noise
\cite{ChateComment}. However, it is not clear how such coupling
intervenes in the onset of collective order in the self-propelled
model, nor how it can change the phase transition from continuous to
discontinuous. To explore this idea, and more importantly, to
determine whether or not the results obtained for the network models
presented here can be applied to the self-propelled model, further
analysis of the relation between the dynamics of a swarming system and
the connectivity of its corresponding network representation is
required.

\section*{Acknowledgments}
This work was supported by the National Science Foundation under Grant
No. DMS-0507745 (C.H.), by CONACyT Grant P47836-F and by PAPIIT-UNAM Grant
IN112407-3 (M.A. and H.L.). J.A.P. acknowledges CONACyT for a
M.Sc. scholarship.

\appendix

\section{Analytic results for the Majority Voter Model}
\label{sec:appendixA}

In this appendix we present the analytic computation leading to
Eqs.~\eqref{eqs:MappingVoter} and \eqref{eq:discontVoter}. Let us
start by defining the quantities

\begin{subequations}
\begin{eqnarray}
 \chi_n(t) &=& 4K\eta_1\xi_n(t),\label{eq:noise}\\ 	
 u_n(t) &=& \sum_{j=1}^K v_{n_j}(t), \label{eq:sumSpins}\\
 s_n(t) &=& u_n(t) + \chi_n(t).\label{eq:sumTotal}
\end{eqnarray}
\end{subequations}
With these definitions, the dynamical interaction rule
Eq.~\eqref{eq:voterGeneral} can be rewritten in the equivalent
form
\begin{equation}
v_n(t+1)=\left\{
\begin{array}{rcc}
 \mbox{Sign}\left[s_n(t)\right] & \mbox{with prob.} & 1-\eta_2\\
& & \\
-\mbox{Sign}\left[s_n(t)\right] & \mbox{with prob.} & \eta_2
\end{array}
\right.
\label{eq:ApVoterGeneral}
\end{equation}

Let $\phi_n(t)$ and $P_n^+(t)$ be the probabilities that $v_n(t)=1$
and $s_n(t) > 0$, respectively. From Eq.~\eqref{eq:ApVoterGeneral} it
is clear that $\phi_n(t)$ and $P_n^+(t)$ are related through
\begin{equation}
 \phi_n(t) = (1-\eta_2)P_n^+(t) + (1-P_n^+(t))\eta_2.
\label{eq:masterVoter}
\end{equation}
The first term on the right-hand side of the above equation accounts
for the case in which $s_n(t)>0$ and the Sign function is
evaluated. The second term takes into account the case in which
$s_n(t)<0$ and the $-$Sign function is evaluated. In both of these
cases $v_n(t+1)$ will be positive.

Now, we introduce the mean-field assumptions used in this computation: 
\begin{itemize}
\item The probabilities $\phi_n(t)$ and $P_n^+(t)$ are the same for
all the elements in the network. This assumption states that all the
network elements are statistically equivalent. Thus, from now on we
will drop off the subscript ``$n$'' in these quantities and use the
notation $\phi_n(t) = \phi(t)$ and $P_n^+(t) = P^+(t)$.

\item That the network elements $v_n$ are statistically
independent implies that the sum $u_n(t)$ defined in
Eq.~\eqref{eq:sumSpins} can be considered as the sum of $K$
independent and identically distributed random variables.
\end{itemize}

Note that the order parameter $\psi(t)=\langle v_n(t)\rangle$ can be
expressed in terms the probability $\phi(t)$ as
\[
 \psi(t) = \langle v_n(t)\rangle = (1)\phi(t) + (-1)(1-\phi(t)) = 2\phi(t)-1,
\]
from which we obtain
\begin{equation}
 \phi(t) = \frac{1+\psi(t)}{2}.
\label{eq:psiphi}
\end{equation}
Eq.\eqref{eq:masterVoter} can then be written in terms of the order
parameter $\psi(t)$ as
\begin{equation}
 \psi(t+1) = 2(1-2\eta_2)P^+(t) + 2\eta_2 -1.
\label{eq:masterVoterPsi}
\end{equation}

The next step in this calculation consists in expressing $P^+(t)$ as a
function of $\phi(t)$, or equivalently, as a function of $\psi(t)$. To
this end, let $P_v(x,t)$, $P_u(x,t)$, $P_s(x,t)$, and $P_\chi(x)$ be
the probability density functions of the random variables $v_n(t)$,
$u_n(t)$, $s_n(t)$ and $\chi_n(t)$, respectively. (Note that
$P_\chi(x)$, the probability density function of the noise, is
independent of time.)  From Eq.~\eqref{eq:sumTotal} it follows that
\begin{equation}
 P_s(x,t) = [P_u*P_\chi](x,t),
\label{eq:PuConvPchiVoter}
\end{equation}
where ``$*$'' denotes a convolution. Analogously, since $u_n(t)$ is
the sum of the $K$ independent random variables
$v_{n_1}(t),\dots,v_{n_K}(t)$, all equally distributed with the PDF
$P_v(x,t)$, then $P_u(x,t)$ is the $K$-fold convolution of $P_v(x,t)$
with itself, and therefore, $P_s(x,t)$ becomes
\begin{equation}
 P_s(x,t) = [\underbrace{P_v*\cdots*P_v}_{K\ times}*P_\chi](x,t).
\label{eq:PsConvolutionVoter}
\end{equation}
It is convenient to transform the above equation to Fourier space (in
the variable $x$). Denoting as $\hat{P}_v(\lambda,t)$,
$\hat{P}_u(\lambda,t)$, $\hat{P}_s(\lambda,t)$ and
$\hat{P}_\chi(\lambda)$ the Fourier transforms of the corresponding
PDF's, we obtain
\begin{equation}
 \hat{P}_s(\lambda,t) = \left[\hat{P}_v(\lambda,t)\right]^K \hat{P}_\chi(\lambda,t).
\label{eq:PsFourierVoter}
\end{equation}

We have now to consider two cases separately, the case for which
$K<\infty$, and the case $K\to\infty$.

\subsection{Case 1: $K<\infty$}

By definition of $\phi(t)$, we have that 
\[
P_v(x,t) = \phi(t)\delta(x-1) + (1-\phi(t))\delta(x+1),
\]
where $\delta(x)$ is the Dirac delta function. Therefore \footnote{We
are using the definition $\hat{f}(\lambda) = \int_{-\infty}^\infty
f(x)e^{-i\lambda x}dx$, $f(x) = (2\pi)^{-1}\int_{-\infty}^\infty
\hat{f}(\lambda)e^{-i\lambda x}d\lambda$ for the Fourier transform and
its inverse.},
\[
 \hat{P}_v(\lambda,t) =\phi(t)e^{-i\lambda}+(1-\phi(t))e^{i\lambda}. 
\]
The above equation can be written in terms of $\psi(t)$ by using the
relationship given in Eq.~\eqref{eq:psiphi}, which gives
\begin{equation}
 \hat{P}_v(\lambda,t) = \cos\lambda - i \psi(t)\sin\lambda.
\label{eq:PvFourierVoter}
\end{equation}

On the other hand, $\chi_n(t)$ is a random variable uniformly
distributed in the interval $[-4K\eta_1,4K\eta_1]$, from which it
follows that
\begin{equation}
\hat{P}_\chi(\lambda) = \frac{1}{4K\eta_1\lambda}\sin{4K\eta_1\lambda}.
\label{eq:PnoiseVoter} 
\end{equation}
Substituting into Eq.~\eqref{eq:PsFourierVoter} the results given in
Eqs.~\eqref{eq:PvFourierVoter} and \eqref{eq:PnoiseVoter} we obtain
\begin{eqnarray}
 \hat{P}_s(\lambda,t) &=& \left[\cos\lambda - i
\psi(t)\sin\lambda\right]^K
\frac{\sin(4K\eta_1\lambda)}{4K\eta_1}\nonumber\\ &=& \sum_{m=0}^K
(-i)^m\binom{K}{m}
[\cos\lambda]^{K-m}[\psi(t)\sin\lambda]^{m}\nonumber \\
&&\times\frac{\sin(4K\eta_1\lambda)}{4K\eta_1\lambda}.
\label{eq:PsFourierBinomVoter}
\end{eqnarray}

Since $P^+(t)$ is the probability that $s_n(t)>0$, then
\[
P^+(t)=\int_0^\infty P_s(x,t)dx.
\]
Therefore, taking the inverse Fourier transform of
Eq.~\eqref{eq:PsFourierBinomVoter} and integrating the result from 0
to $\infty$ we obtain
\begin{equation}
 P^+(t) = \frac{1}{2}\sum_{m=0}^K \beta_m^K(\eta_1) [\psi(t)]^m,
\label{eq:P+1}
\end{equation}
where the coefficients $\beta_m^K(\eta_1)$ are given by
\begin{eqnarray}
\beta_m^K(\eta_1)&=&\binom{K}{m}\frac{(-i)^m}{\pi4 K\eta_1}\int_0^\infty\int_{-\infty}^\infty [\cos\lambda]^{K-m}[\sin\lambda]^{m}\nonumber\\
&&\times\frac{\sin(4K\eta_1\lambda)}{\lambda}e^{i\lambda x} d\lambda dx. 
\label{eq:ApBeta1}
\end{eqnarray}

Although it is not obvious from the above expression, it happens that
$\beta_0^K(\eta_1)=1$. To show that this is indeed the case, let us
define the function $\hat{G}(\lambda)$ as
\[
 \hat{G}(\lambda)= [\cos\lambda]^K\frac{\sin(4K\eta_1\lambda)}{4K\eta_1\lambda}.
\]
Note that $\hat{G}(\lambda)$ is a symmetric function and that
$\hat{G}(0)=1$. With this definition, the coefficient
$\beta_0^K(\eta_1)$ can be written as
\begin{eqnarray*}
\beta_0^K(\eta_1)&=& 2\int_0^\infty\frac{1}{2\pi}\int_{-\infty}^{\infty}\hat{G}(\lambda)e^{i\lambda x}d\lambda dx\\
&=&2\int_0^\infty G(x)dx,
\end{eqnarray*}
where $G(x)$ is the inverse Fourier transform of
$\hat{G}(\lambda)$. Since $\hat{G}(\lambda)$ is symmetric, then $G(x)$
is also symmetric and therefore $\int_{-\infty}^\infty G(x)dx =
2\int_{0}^\infty G(x)dx$. Additionally, since $\hat{G}(0)=1$ then
$\int_{-\infty}^\infty G(x)dx = 1$, from which it follows that
$\beta_0^K(\eta_1)=1$.

For $m\geq1$ we can exchange the order of integration in
Eq.~\eqref{eq:ApBeta1} by multiplying the integrand by $e^{-\epsilon
x}$. After performing the integral over $x$ and then taking the limit
$\epsilon\to0$ we obtain
\begin{eqnarray}
\beta_m^K(\eta_1)&=&\binom{K}{m}\frac{(-i)^{m-1}}{\pi4 K\eta_1}\nonumber\\
&\times&\int_{-\infty}^\infty [\cos\lambda]^{K-m}[\sin\lambda]^{m}
\frac{\sin(4K\eta_1\lambda)}{\lambda^2} d\lambda.
\end{eqnarray}

Using the fact that $\beta_0^K(\eta_1)=1$, Eq.~\eqref{eq:P+1} can be
written as
\[
 P^+(t) = \frac{1}{2}\left(1 +\sum_{m=1}^K\beta_m^K(\eta_1)[\psi(t)]^m\right).
\label{eq:ApUltima}
\] 
Finally, substituting the above result into
Eq.~\eqref{eq:masterVoterPsi} we obtain
\begin{equation}
 \psi(t+1)=(1-2\eta_2)\sum_{m=1}^K\beta_m^K(\eta_1)[\psi(t)]^m.
\label{eq:ApFinalPsi}
\end{equation}
%

\subsection{Case 2: $K\to\infty$}

By definition [see Eq.~\eqref{eq:sumSpins}] $u_n(t)$ is the sum of $K$
independent and identically distributed variables, each with average
$\psi(t)$ and variance $1-[\psi(t)]^2$. Therefore, for very large
values of $K$ the Central Limit Theorem allows us to approximate
$P_u(x,t)$ by a Gaussian with average $K\psi(t)$ and variance
$\sigma^2=K(1-[\psi(t)]^2)$
\[
 P_u(x,t) \approx 
\frac{\exp\left(-\frac{(x-K\psi(t))^2}{2K(1-[\psi(t)]^2)}\right)}
{\sqrt{2\pi K(1-[\psi(t)]^2)}}. 
\]
With this approximation, Eq.~\eqref{eq:PuConvPchiVoter} becomes
\begin{eqnarray*}
 P_s(x,t)&=&\frac{1}{8K\eta_1\sqrt{2\pi K(1-[\psi(t)]^2)}}\\
&\times& \int_{-4K\eta_1}^{4K\eta_1}
\exp\left(-\frac{(x-y-K\psi(t))^2}{2K(1-[\psi(t)]^2)}\right)dy,
\end{eqnarray*}
where we have used the fact that $P_\chi(x)$ is a constant normalized
function defined in the interval $x\in[-4K\eta_1,4K\eta_1]$. From the
above expression we can obtain $P^+(t)$ by integrating $P_s(x,t)$ from
0 to $\infty$:
\begin{eqnarray*}
 P^+(t)&=&\frac{1}{8K\eta_1\sqrt{2\pi K(1-[\psi(t)]^2)}}\\
&\times& \int_0^\infty \int_{-4K\eta_1}^{4K\eta_1}
\exp\left(-\frac{(x-y-K\psi(t))^2}{2K(1-[\psi(t)]^2)}\right)dy dx.
\end{eqnarray*}
Performing the change of variable $x = Kx'$, $y = Ky'$, the above
expression transforms into
\begin{eqnarray*}
 P^+(t)&=&\frac{1}{8\eta_1\sqrt{\frac{2\pi}{K}(1-[\psi(t)]^2)}}\\
&\times& \int_0^\infty \int_{-4\eta_1}^{4\eta_1}
\exp\left(-\frac{(x'-y'-\psi(t))^2}{\frac{2}{K}(1-[\psi(t)]^2)}\right)dy' dx'.
\end{eqnarray*}
In the limit $K\to\infty$ we have
\[
\lim_{K\to\infty} \frac{\exp\left(-\frac{(x'-y'-\psi(t))^2}{\frac{2}{K}(1-[\psi(t)]^2)}\right)}{\sqrt{\frac{2\pi}{K}(1-[\psi(t)]^2)}}
= \delta\left(x'-y'-\psi(t)\right),
\]
where $\delta(\cdot)$ is the Dirac delta function. From the last two
equations above it follows that, in the limit $K\to\infty$, the
probability $P^+(t)$ acquires the simpler form
\[
 P^+(t) = \frac{1}{8\eta_1}\int_0^\infty \int_{-4\eta_1}^{4\eta_1} \delta\left(x'-y'-\psi(t)\right)dx' dy'.
\]
After performing the integrals in the above expression we obtain
\[
 P^+(t) =
\left\{ 
\begin{array}{llr}
0 & \mbox{ if } & \psi(t) < -4\eta_1\\ 
&&\\
\frac{1}{8\eta_1}(\psi(t) + 4\eta_2) & \mbox{ if } & |\psi(t)|\leq 4\eta_1\\
&&\\
1 & \mbox{ if } & \psi(t) > 4\eta_1
\end{array}
\right.
\]
This last result combined with Eq.~\eqref{eq:masterVoterPsi} give
Eq.~\eqref{eq:discontVoter} of the main text.

\section{Analytic computation of the phase transition for the
  Vectorial network model}
\label{sec:appendixB}

The dynamics of the Vectorial network model (VNM) are given by the
interaction rule
\begin{equation}
\theta_m(t+1)=\mbox{Angle}\left[\sum_{j=1}^K \vec{v}_{m_j}(t) + K\eta_1 e^{i\xi_m(t)}\right]
+ \eta_2\zeta_m(t),
\label{eq:chaterule}
\end{equation}
where $\{\vec{v}_{m_j}\}_{j=1}^K$ are the $K$ inputs of $\vec{v}_m$,
and $\xi_m(t)$ and $\zeta_m(t)$ are independent random variables
uniformly distributed in the interval $[-\pi,\pi]$ and
$0\leq\eta_1,\eta_2\leq1$. Let us define the extrinsic noise vector
$\vec{n}_e$ and the intrinsic noise $n_i$ as
\begin{eqnarray*}
 \vec{n}_e &=& K\eta_1  e^{i\xi(t)}, \\
 n_i &=& \eta_2 \zeta.
\end{eqnarray*}
We will denote as $P_{\vec{n}_e}(r,\xi)$ the PDF (in polar
coordinates) of the extrinsic noise $\vec{n}_e$ and as
$P_{n_i}(\zeta)$ the PDF of the intrinsic noise $n_i$. As for the
majority voter model, it is convenient to define the quantities
\begin{eqnarray}
\vec{u}_m(t) &=& \sum_{j=1}^K\vec{v}_{m_j}(t), \\
\vec{s}_m(t) &=& \vec{u}_m(t) + \vec{n}_e(t).
\end{eqnarray}
Let $(v_m,\theta_m)$, $(u_m,\beta_m)$ and $(s_m,\theta'_m)$ be the
polar coordinates of the vectors $\vec{v}_m(t)$, $\vec{u}_m(t)$, and
$\vec{s}_m(t)$, respectively. We will denote as
$P_{\vec{v}_m}(v_m,\theta_m;t)$, $P_{\vec{u_m}}(u_m,\beta_m;t)$, and
$P_{\vec{s}_m}(s_m,\theta'_m;t)$ the PDF's of these three vectors,
respectively. In Table~\ref{tab:1} we summarize the relevant
quantities appearing in this calculation.

\begin{table}[t]
 \begin{tabular}{|l|c|c|c|}
\hline
\textbf{Scalar} & \textbf{Amplitude} & \textbf{PDF} & \textbf{Fourier Transform}\\
\hline
$n_i = \eta_2\zeta(t)$ & $\eta_2$ & $P_{n_i}(\zeta)$ & $\hat{p}_m$\\
\hline
\hline
\textbf{Vector} & \textbf{Polar Coor.} &  \textbf{PDF} & \textbf{Fourier Transform}\\
\hline
$\vec{n}_e(t) = K \eta_1 e^{i\xi(t)}$ & $(r,\xi)$ & $P_{\vec{n}_e}(r,\xi)$ & 
$\hat{P}_{\vec{n}_e}(\lambda,\gamma)$\\
\hline
$\vec{v}_m(t) = e^{i\theta_m(t)}$ & $(v,\theta)$ & $P_{\vec{v}}(v,\theta;t)$ &
$\hat{P}_{\vec{v}}(\lambda,\gamma;t)$\\
\hline	
$\vec{u}_m(t) = \sum_{j=1}^K\vec{v}_{m_j}(t)$ & $(u,\beta)$ & $P_{\vec{u}}(u,\beta;t)$ &
$\hat{P}_{\vec{u}}(\lambda,\gamma;t)$\\
\hline
$\vec{s}_m(t) = \vec{u}_m(t) + \vec{\eta}(t)$ & $(s,\theta')$ & $P_{\vec{s}}(s,\theta';t)$  &
$\hat{P}_{\vec{s}}(\lambda,\gamma;t)$\\
\hline
 \end{tabular}
\label{tab:1}
\caption{Notation guide for the different quantities involved in the
calculation of the phase transition of the VNM. We have omitted the
subscript $m$ in the PDF's since we assume that all the network
elements $\vec{v}_m$ are statistically equivalent.}
\end{table}

Note that neither $P_{\vec{n}_e}(r,\xi)$ nor $P_{n_i}(\zeta)$ depend
on time or on the subscript $m$ of $\vec{v}_m$, whereas the PDF's of
$\vec{v}_m(t)$, $\vec{u}_m(t)$, and $\vec{s}_m(t)$ depend on both time
and the subscript $m$. However, in a mean-field approximation, we can
assume that all the vectors $\vec{v}_m$ are \emph{statistically
equivalent} and \emph{statistically independent}. In this case, the
functions $P_{\vec{v}_m}(v_m,\theta_m;t)$,
$P_{\vec{u_m}}(u_m,\beta_m;t)$, and $P_{\vec{s}_m}(s_m,\theta'_m;t)$
are site independent, (i.e., the same for all the vectors in the
network) and the subscript $m$ can be omitted. From now on we will
assume that the conditions for the validity of the mean-field
approximation (statistical equivalence and independence) apply.

\subsection{The order parameter}

To measure the amount of order in the system, we define the
instantaneous order parameter $\psi(t)$ as
\begin{equation}
\psi(t) = \left|\frac{1}{N}\sum_{m=1}^\infty \vec{v}_m(t)\right| = \left|\left\langle \vec{v}(t)\right\rangle\right|,
\end{equation}
where we have defined $\left\langle \vec{v}(t)\right\rangle =
\frac{1}{N}\sum_{m=1}^\infty \vec{v}_m(t)$. Under the mean-field
assumption, all the vectors $\vec{v}_m$ are equally distributed with
the common probability distribution $P_{\vec{v}}(v,\theta;t)$. Then
$\left\langle \vec{v}(t)\right\rangle$ can be computed as follows.

Let $\hat{P}_{\vec{v}}(\lambda,\gamma;t)$ be the Fourier transform (in
polar coordinates) of $P_{\vec{v}}(v,\theta;t)$. The variables
$\lambda$ and $\gamma$ are the Fourier conjugates of the variables $v$
and $\theta$, respectively. A cumulant expansion of
$\hat{P}_{\vec{v}}(\lambda,\gamma;t)$ up to the first order gives
\begin{equation}
\hat{P}_{\vec{v}}(\lambda,\gamma;t) \approx
1 - i\langle\vec{v}(t)\rangle\cdot\vec{\lambda} + \cdots,
\label{eq:param1}
\end{equation}
where $\vec{\lambda}$ is the vector in Fourier space whose polar
 coordinates are $(\lambda,\gamma)$. Denoting as $\alpha$ the angle
 between $\langle \vec{v}(t)\rangle$ and $\vec{\lambda}$, and using
 the fact that $\psi(t) = \left|\langle \vec{v}(t)\rangle\right|$,
 Eq.~(\ref{eq:param1}) can be written as 
\begin{equation}
\hat{P}_{\vec{v}}(\lambda,\gamma;t) \approx
1 - i\psi(t)\lambda\cos\alpha + \cdots.
\label{eq:radialMoment}
\end{equation}
Thus, a first order cumulant expansion of
$\hat{P}_{\vec{v}}(\lambda,\gamma;t)$ directly gives us the order
parameter $\psi(t)$. The objective of the calculation is to find a
recurrence relation in time for $\hat{P}_{\vec{v}}(\lambda,\gamma;t)$
based on Eq.~(\ref{eq:chaterule}). From this recurrence relation we
will obtain the dynamical mapping that determines the temporal
evolution of $\psi(t)$.

\subsection{Recurrence relation for $\hat{P}_{\vec{v}}(\lambda,\gamma;t)$}

Note first that, since $|\vec{v}_m|=1$ for all $m$, then
$P_{\vec{v}}(v,\theta;t)$ can be written as
\begin{equation}
 P_{\vec{v}}(v,\theta;t) = \frac{\delta(v-1)}{v}P_{\theta}(\theta;t),
\label{eq:Pv}
\end{equation}
where $P_{\theta}(\theta;t)$ is the PDF of the angle $\theta(t)$ of
$\vec{v}(t)$. From Eq.~(\ref{eq:chaterule}) it follows that
$P_{\theta}(\theta;t)$, $P_{\vec{s}}(s,\theta;t)$ and $P_{n_i}(\zeta)$
are related through
\begin{equation}
 P_{\theta}(\theta;t+1) = \int_{-\pi}^{\pi}\left[
\int_0^\infty s P_{\vec{s}}(s,\theta-\zeta;t)ds\right]P_{n_i}(\zeta)d\zeta.
\label{eq:mainIntegral}
\end{equation}
Since $\vec{s}(t) = \sum_{j=1}^K\vec{v}_{m_j}(t) + \vec{n}_e(t)$, and
each of the vectors $\vec{v}_{m_j}(t)$ is distributed with the PDF
$P_{\vec{v}}(v,\theta;t)$, it is clear that $P_{\vec{s}}(s,\theta;t)$
depends on $P_{\theta}(\theta;t)$. Therefore,
Eq.~(\ref{eq:mainIntegral}) is a recurrence relation in time for
$P_{\theta}(\theta;t)$. This recurrent relation is best solved in
Fourier space. Denoting as $\hat{P}_{\vec{s}}(\lambda,\gamma;t)$ the
Fourier transform of $P_{\vec{s}}(s,\theta;t)$, the above equation can
be written as
\begin{eqnarray}
 P_{\theta}(\theta;t+1) &=& \frac{1}{(2\pi)^2}
\int_{-\pi}^{\pi}d\zeta\int_0^\infty sds \int_0^\infty \lambda d\lambda 
\int_0^{2\pi}d\gamma  \nonumber \\
&\times&\hat{P}_{\vec{s}}(\lambda,\gamma-\zeta;t) P_{n_i}(\zeta) e^{is\lambda \cos(\gamma-\theta)}.
\label{eq:mainIntegralFourier}
\end{eqnarray}
Since $P_{\theta}(\theta;t)$, $\hat{P}_{\vec{s}}(\lambda,\gamma;t)$
and $P_{n_i}(\zeta)$ are periodic functions of their angular arguments
($\theta$, $\gamma$ and $\zeta$ respectively), we can expand these
functions in Fourier series as
\begin{eqnarray}
 P_{\theta}(\theta;t) &=& \sum_{m=-\infty}^\infty \phi_m(t)e^{im\theta},
\label{eq:Pthetaphi}\\
 \hat{P}_{\vec{s}}(\lambda,\gamma-\zeta;t) &=& \sum_{m=-\infty}^\infty \chi_m(\lambda;t)e^{im(\gamma-\zeta)},
\label{eq:Pschi}\\
P_{n_i}(\zeta) &=& \sum_{m=-\infty}^\infty \hat{p}_me^{-im\zeta},
\label{eq:PniSeries}
\end{eqnarray}
where $\phi_m(t)$, $\chi_m(\lambda;t)$ and $\hat{p}_m$ are given by
\begin{eqnarray}
 \phi_m(t) &=& \frac{1}{2\pi}\int_{-\pi}^{\pi} P_{\theta}(\theta;t) e^{-im\theta}d\theta,\\
 \chi_m(\lambda;t) &=&  \frac{1}{2\pi}\int_{-\pi}^{\pi} \hat{P}_{\vec{s}}(\lambda,\gamma;t) e^{-im\gamma}d\gamma, 
\label{eq:chimPs}\\
\hat{p}_m &=& \frac{1}{2\pi}\int_{-\pi}^{\pi} P_{n_i}(\zeta) e^{-im\zeta}d\zeta.
\label{eq:PniFourier}
\end{eqnarray}
Substituting Eqs.\eqref{eq:Pthetaphi} and \eqref{eq:Pschi} into
Eq.~(\ref{eq:mainIntegralFourier}), carrying out the integration over
$\zeta$ and taking into account Eq.\eqref{eq:PniFourier} we obtain
\begin{eqnarray*}
 &&\sum_{m=-\infty}^\infty \phi_m(t+1)e^{im\theta} \\
 &=&\sum_{m=-\infty}^\infty (i)^m \hat{p}_m \int_0^\infty s ds \int_0^\infty \lambda d\lambda
\chi_m(\lambda;t) J_m(s\lambda) e^{im\theta}, 
\end{eqnarray*}
where we have used the integral representation of the Bessel function
$J_m(x) = \frac{(-i)^m}{2\pi} \int_0^{2\pi} e^{i(mz + x\cos z)}
dz$. It follows from the last expression that
\begin{equation}
 \phi_m(t+1) = i^m \hat{p}_m\int_0^\infty s ds \int_0^\infty \lambda d\lambda
\chi_m(\lambda;t)J_m(s\lambda).
\label{eq:phichi1}
\end{equation}

Exchanging the order of integration in the last expression, and using
the identity
\[
\int_0^\infty sJ_m(\lambda s)ds = \frac{\delta(\lambda)}{\lambda}\delta_{m,0} +
\frac{m}{\lambda^2},
\]
where $\delta(\lambda)$ and $\delta_{m,0}$ are the Dirac and Kronecker
delta functions, respectively, Eq.~(\ref{eq:phichi1}) becomes
\begin{equation}
\phi_m(t+1) = (i)^m \hat{p}_m\left(
\delta_{m,0}\chi_m(0;t) + m\int_0^\infty\chi_m(\lambda;t)\frac{d\lambda}{\lambda}\right).
\label{eq:phichi2}
\end{equation}

Note that Eq.~(\ref{eq:phichi2}) is a consequence of the recurrence
relation given in Eq.~(\ref{eq:mainIntegral}), which in turn follows
directly from the dynamic interaction rule
Eq.~(\ref{eq:chaterule}). Now we have to project the probability
distribution function $P_{\vec{s}}(s,\theta;t)$ onto the unit circle
by forcing the vector $\vec{s}(t)$ to have unit length at time $t+1$,
and thus becoming $\vec{v}(t)$. To do so, we take the Fourier
transform of $P_{\vec{v}}(v,\theta;t)$ given in Eq.~(\ref{eq:Pv}),
which when evaluated at time $t+1$ gives
\begin{equation}
\hat{P}_{\vec{v}}(\lambda,\gamma;t+1)=\int_0^{2\pi}P_{\theta}(\theta;t+1)e^{-i\lambda\cos(\theta-\gamma)}d\theta.
\end{equation}
Substituting into the above equation the form of
$P_{\theta}(\theta;t)$ given in Eq.~(\ref{eq:Pthetaphi}) (evaluated at
$t+1$), we obtain
\begin{equation}
\hat{P}_{\vec{v}}(\lambda,\gamma;t+1) = 2\pi\sum_{m=-\infty}^\infty
(-i)^m\phi_m(t+1)  J_m(\lambda) e^{im\gamma},
\end{equation}
where we have used the integral representation of the Bessel function
$J_m(\lambda)=\frac{(i)^m}{2\pi} \int_0^{2\pi}e^{i(mz-\lambda\cos
z)}dz$. Now we use the value of $\phi_m(t+1)$ given in
Eq.~(\ref{eq:phichi2}), which leads to
\begin{eqnarray}
 \hat{P}_{\vec{v}}(\lambda,\gamma;t+1) &=& 2\pi\hat{p}_0 J_0(\lambda) 
+ \sum_{m=-\infty}^\infty 2\pi m \hat{p}_m
J_m(\lambda) e^{im\lambda} \nonumber\\
&\times& \int_0^\infty \chi_m(\lambda';t)\frac{d\lambda'}{\lambda'}.
\label{eq:quasifinal}
\end{eqnarray}
To complete the projection of $\hat{P}_{\vec{v}}(\lambda,\gamma;t)$
onto the unit circle in a closed form, it only remains to find
$\chi_m(t)$ as a function of
$\hat{P}_{\vec{v}}(\lambda,\gamma;t)$. Since $s(t) =
\sum_{j=1}^K\vec{v}_{m_j}(t) + \vec{n}_e(t)$, and we are assuming that
all the $\vec{v}_j$ are statistically independent, then
\begin{equation}
 \hat{P}_{\vec{s}}(\lambda,\gamma;t) = \left[\hat{P}_{\vec{v}}(\lambda,\gamma;t)\right]^K 
 \hat{P}_{\vec{n}_e}(\lambda,\gamma),
\end{equation}
where $\hat{P}_{\vec{n}_e}(\lambda,\gamma)$ is the Fourier transform
of the PDF of the noise vector $\vec{n}_e=K\eta_1 e^{i\xi}$. Since
$\xi$ is uniformly distributed in the interval $[0,2\pi]$, it follows
that $\hat{P}_{\vec{n}_e}(\lambda,\gamma) =
J_0(K\eta\lambda)$. Therefore, we obtain
\begin{equation}
 \hat{P}_{\vec{s}}(\lambda,\gamma;t) = \left[\hat{P}_{\vec{v}}(\lambda,\gamma;t)\right]^K 
 J_0(K\eta\lambda).
\end{equation}
Substituting the above expression into Eq.~(\ref{eq:chimPs}) we obtain
\begin{equation}
 \chi_m(\lambda;t) = \frac{1}{2\pi} J_0(K\eta\lambda) \int_0^{2\pi}
\left[\hat{P}_{\vec{v}}(\lambda,\gamma;t)\right]^K 
  e^{-im\gamma}d\gamma.
\end{equation}
Finally, combining this result with Eq.~(\ref{eq:quasifinal}), we
obtain the desired recurrence relation for
$\hat{P}_{\vec{v}}(\lambda,\gamma;t)$:
\begin{eqnarray}
 && \hat{P}_{\vec{v}}(\lambda,\gamma;t+1) = 2\pi\hat{p}_0 J_0(\lambda) + \sum_{m=-\infty}^\infty
 m \hat{p}_m J_m(\lambda)e^{im\gamma}\nonumber\\
 &\times& \int_{0}^\infty \frac{d\lambda'}{\lambda'}J_0(K\eta\lambda')\int_{0}^{2\pi}d\gamma' 
 \left[\hat{P}_{\vec{v}}(\lambda',\gamma';t)\right]^K 
  e^{-im\gamma'}.
\label{eq:recurrenceFinal}
\end{eqnarray}

\subsection{Dynamical mapping for $K\to\infty$}

Eq.~\eqref{eq:recurrenceFinal} is a complicated recurrence relation
the exact solution to which is way out of our hands. However, for
large values of $K$, namely, for a large number of interactions per
particle, we can use the Central Limit Theorem to approximate
$\left[\hat{P}_{\vec{v}}(\lambda,\gamma;t)\right]^K $ as
\[
 \left[\hat{P}_{\vec{v}}(\lambda,\gamma;t)\right]^K \approx
  \exp\left\{-iK\langle \vec{v}(t)\rangle\cdot\vec{\lambda}-
 \frac{K}{2}\vec{\lambda}\cdot{\mathbf C}(t)\cdot \vec{\lambda}\right\},
\]
where $\langle \vec{v}(t)\rangle$ and ${\mathbf C}(t)$ are the first
moment and covariance matrix of $P_{\vec{v}}(v,\theta;t)$,
respectively, and $\vec{\lambda}$ is the vector in Fourier space with
polar coordinates $(\lambda,\gamma)$. With this approximation,
Eq.~(\ref{eq:recurrenceFinal}) becomes
\begin{eqnarray*}
&& \hat{P}_{\vec{v}}(\lambda,\gamma;t+1) = 2\pi \hat{p}_0 J_0(\lambda) + \sum_{m=-\infty}^\infty
 m \hat{p}_m J_m(\lambda)e^{im\gamma} \\
 &\times& \int_{0}^\infty \frac{d\lambda'}{\lambda'}J_0(K\eta\lambda')\int_{0}^{2\pi}d\gamma' 
 e^{-iK\langle \vec{v}(t)\rangle\cdot\vec{\lambda'}-
 \frac{K}{2}\vec{\lambda'}\cdot{\mathbf C}(t)\cdot \vec{\lambda'}}
  e^{-im\gamma'}.
\end{eqnarray*}
Making the change of variable $\vec{x} = K\vec{\lambda'}$ in the above
expression, we obtain
\begin{eqnarray*}
&& \hat{P}_{\vec{v}}(\lambda,\gamma;t+1) = 2\pi\hat{p}_0 J_0(\lambda) + \sum_{m=-\infty}^\infty
 m \hat{p}_m J_m(\lambda)e^{im\gamma} \\
 &\times& \int_{0}^\infty \frac{dx}{x}J_0(\eta x)\int_{0}^{2\pi}d\gamma' 
 e^{-i\langle \vec{v}(t)\rangle\cdot\vec{x}-
 \frac{1}{2K}\vec{x}\cdot{\mathbf C}(t)\cdot \vec{x}}
  e^{-im\gamma'}.
\end{eqnarray*}
We can go a step further in the large-$K$ approximation and neglect
the term $\frac{1}{2K}\vec{x}\cdot{\mathbf C}(t)\cdot \vec{x}$
appearing in the exponent inside the integral of the last expression,
which gives
\begin{eqnarray}
 \hat{P}_{\vec{v}}(\lambda,\gamma;t&+&1) = 2\pi \hat{p}_0 J_0(\lambda) + \sum_{m=-\infty}^\infty
 m\hat{p}_m J_m(\lambda)e^{im\gamma} \nonumber\\
 &\times& \int_{0}^\infty \frac{dx}{x}J_0(\eta x)\int_{0}^{2\pi}d\gamma' 
 e^{-i\left[ m\gamma' + \langle \vec{v}(t)\rangle\cdot\vec{x}\right]}.
\label{eq:yacasi}
\end{eqnarray}

Now, we can write $\langle \vec{v}(t)\rangle\cdot\vec{x} =
\left|\langle \vec{v}(t)\rangle\right| x \cos(\gamma' - \alpha)$,
where $\gamma'$ and $\alpha$ are the angles in Fourier space of
$\vec{x}$ and $\langle\vec{v}(t)\rangle$, respectively. The second
integral on the right-hand side of Eq.~(\ref{eq:yacasi}) becomes
\begin{eqnarray*}
 \int_{0}^{2\pi} e^{-i\left[m\gamma'+ \langle \vec{v}(t)\rangle\cdot\vec{x}\right]}d\gamma' &=&
\int_{0}^{2\pi}e^{-i\left[m\gamma'+ |\langle \vec{v}(t)\rangle|x\cos(\gamma'-\alpha)\right]}d\gamma' \\
&=& e^{-im\alpha}\int_{0}^{2\pi}e^{-i\left[m\tau+ |\langle \vec{v}(t)\rangle|x\cos \tau\right]}d\tau \\
&=& e^{-im\alpha}2\pi(-i)^m J_m\left(|\langle \vec{v}(t)\rangle|x\right) \\
&=& e^{-im\alpha}2\pi(-i)^m J_m\left(\psi(t)x\right),
\end{eqnarray*}
where we have used the fact that $\psi(t)=|\langle
\vec{v}(t)\rangle|$. Substituting this result into
Eq.~(\ref{eq:yacasi}), we obtain
\begin{eqnarray}
 \hat{P}_{\vec{v}}(\lambda,\gamma;t+1) &=& 2\pi \hat{p}_0 J_0(\lambda) + \sum_{m=-\infty}^\infty
 2\pi (-i)^m  m\hat{p}_m J_m(\lambda) \nonumber\\
 &\times& e^{im(\gamma-\alpha)} \int_{0}^\infty \frac{dx}{x}J_0(\eta x) J_m\left(\psi(t) x\right).
\label{eq:yamero1}
\end{eqnarray}

Now, recalling that $P_{n_i}(\zeta)$ is a constant normalized function
in the interval $[-\pi\eta_2,\pi\eta_2]$, with $0\leq\eta_2\leq1$, its
Fourier transform $\hat{p}_m$ is given by
\begin{equation}
 \hat{p}_m = \frac{\sin(\pi m\eta_2)}{2\pi^2 m\eta_2}. 
\label{eq:pm}
\end{equation}
Thus, $2\pi\hat{p}_0 = 1$ and Eq.~\eqref{eq:yamero1} can be written as
\begin{eqnarray}
 \hat{P}_{\vec{v}}(\lambda,\gamma;t+1) &=& J_0(\lambda) + \sum_{m=-\infty}^\infty
  (-i)^m  2\pi m\hat{p}_m J_m(\lambda) \nonumber\\
 &\times& e^{im(\gamma-\alpha)} \int_{0}^\infty \frac{dx}{x}J_0(\eta x) J_m\left(\psi(t) x\right).
\label{eq:yamero2}
\end{eqnarray}
Finally, expanding both sides of the above equation up to the first
order in $\lambda$, and recalling Eq.~(\ref{eq:radialMoment}) for
the left-hand side, we obtain the recurrence relation for the order
parameter
\begin{equation}
 \psi(t+1) = 2\pi\hat{p}_1 \int_0^\infty J_0(\eta x) J_1\left(\psi(t) x\right)\frac{dx}{x}.
\label{eq:quasiFinalPsi}
\end{equation}
This is Eq.~\eqref{eq:recur3MainText} of the main text.

\end{document}